\documentclass[a4paper,12pt]{article}

\usepackage{amssymb,amstext,amsmath,bm}
\usepackage{tikz-cd}
\usepackage{cite}

\newcommand{\bra}[1]{{\langle{#1}\vert}}
\newcommand{\ket}[1]{{\vert{#1}\rangle}}
\newcommand{\bracket}[2]{\langle #1 \vert #2 \rangle}

\renewcommand{\Re}{\mathop{\rm Re}\nolimits}

\newcommand{\prirodni}{\ensuremath{\mathbb{N}}}
\newcommand{\realni}{\ensuremath{\mathbb{R}}}
\newcommand{\kompleksni}{\ensuremath{\mathbb{C}}}
\newcommand{\del}{\partial}

\newcommand{\orto}{\bot}

\newcommand{\ds}{\displaystyle}

\newcommand{\itGamma}{ {\varGamma} }                                    
\newcommand{\mitGamma}{ {\itGamma} }

\newcommand{\cC}{{\cal C}}
\newcommand{\cD}{{\cal D}}
\newcommand{\cH}{{\cal H}}

\newcommand{\cM}{{\cal M}}

\newcommand{\cO}{{\cal O}}

\newcommand{\GammaEnt}[2]{ \bm{\mitGamma}^{#1}{}_{#2} }
\newcommand{\gEnt}{\bm{g}}
\newcommand{\DeltagEnt}{{\bm{\Delta g}}}
\newcommand{\TEnt}{{\bm{T}}}
\newcommand{\DeltaTEnt}{{\bm{\Delta T}}}
\newcommand{\PsiEnt}{{\bm{\Psi}}}
\newcommand{\nablaEnt}{\bm{\nabla}\!}

\newcommand{\barg}{h}
\newcommand{\barT}{t}

\begin{document}

\title{\bf Entanglement-induced deviation from the geodesic motion in quantum gravity}

\author{Francisco Pipa$^{a,b}$, Nikola Paunkovi\'c$^{c,d}$ and Marko Vojinovi\'c$^{e}$}

\date{}

\maketitle

\begin{center}
\small {\it
$^a$ Department of Philosophy, University of Kansas, 1445 Jayhawk Blvd., \\ Wescoe Hall, Lawrence, KS 66045-7590, U.S.A. \\
$^b$ Departamento de F\'{i}sica, Instituto Superior T\'{e}cnico, Universidade de Lisboa, \\ Avenida Rovisco Pais 1049-001, Lisboa, Portugal \\
$^c$ Instituto de Telecomunica\c{c}\~oes, Av. Rovisco Pais 1, 1049-001, Lisbon, Portugal \\
$^d$ Departamento de Matem\'{a}tica, Instituto Superior T\'{e}cnico, Universidade de Lisboa, \\ Av. Rovisco Pais 1, 1049-001 Lisboa, Portugal \\
$^e$ Institute of Physics, University of Belgrade, Pregrevica 118, 11080 Belgrade, Serbia \\
}

\bigskip

E-mail: \texttt{franciscosapipa@gmail.com}, \texttt{npaunkov@math.tecnico.ulisboa.pt}, \texttt{vmarko@ipb.ac.rs}
\end{center}

\begin{abstract}
We study the derivation of the effective equation of motion for a pointlike particle in the framework of quantum gravity. Just like the geodesic motion of a classical particle is a consequence of classical field theory coupled to general relativity, we introduce the similar notion of an effective equation of motion, but starting from an abstract quantum gravity description. In the presence of entanglement between gravity and matter, quantum effects give rise to modifications of the geodesic trajectory, primarily as a consequence of the interference between various coherent states of the gravity-matter system. Finally, we discuss the status of the weak equivalence principle in quantum gravity and its possible violation due to the nongeodesic motion.
\end{abstract}

\section{\label{SecIntroduction}Introduction}

The formulation of the theory of quantum gravity (QG) is one of the most fundamental open problems in modern theoretical physics. In models of QG, as in any quantum theory, superpositions of states are allowed. In a tentative ``theory of everything'', which includes both gravity and matter at a fundamental quantum level, superpositions of product gravity-matter states are particularly interesting. Entangled states are highly nonclassical, and as such are especially relevant because they give rise to a drastically different behavior of matter from what one would expect based on classical intuition, as confirmed by numerous examples from the standard quantum mechanics (QM). Therefore, it is interesting to study such states in the context of a QG coupled to matter, in particular the Schr\"odinger cat-like states. Moreover, a recent study \cite{PaunkovicVojinovic2017} suggests that physically allowed states of a gravity-matter system are generically entangled due to gauge invariance, providing additional motivation for our study.

In standard QM, entanglement is generically a consequence of the interaction. Nevertheless, there exist situations which give rise to entanglement even without interaction. For example, the Pauli exclusion principle in the case of identical particles generates entanglement without an interaction, giving rise to an effective force (also called the ``exchange interaction''). We investigate in detail whether an entanglement between gravity and matter could also be described as a certain type of an effective interaction, and if so, what are its aspects and details. In order to study this problem, we analyze the motion of a free test particle in a gravitational field. In general relativity (GR), this motion is described by a geodesic trajectory. However, we show that in the presence of the gravity-matter entanglement, the resulting effective interaction causes a deviation from a classical geodesic trajectory. In particular, we generalize the standard derivation of a geodesic equation from the case of classical gravity to the case of a full QG model, and derive the equation of motion for a particle which contains a non-geodesic term, reflecting the presence of the entanglement-induced effective interaction. The effects we discuss are purely quantum with respect to both gravity and matter, unlike previous studies of quantum matter in classical curved spacetime~\cite{Chowdhury,ZychBrukner2015,Rosi,Longhi}.

As a consequence of the modified equation of motion for a particle, we also discuss the status of the equivalence principle in the context of QG, and a possible violation of its weak flavor.

The paper is organized as follows. Section~\ref{SecGeodesicEquationInGR} is devoted to a review of the derivation of the geodesic equation in classical gravity, particularly in GR. The multipole formalism is employed and the geodesic equation for a particle is derived from the covariant conservation of the stress-energy tensor. In section~\ref{SecGeodesicEquationInQG} we generalize this procedure and derive our main results. Subsection~\ref{SubSecPrelimiaries} contains the general setup, the abstract quantum gravity framework that will be used, and the main assumptions. In subsection~\ref{SubSecCovariantConservationEquation} we discuss the effective covariant conservation equation, which receives a correction to the classical one, due to the quantum gravity effects. In subsection~\ref{SubSecEquationOfMotion} we put everything together and derive our main result --- the effective equation of motion for a point particle, with the leading quantum correction. In subsection~\ref{SubSecConsistency} we discuss the consistency of the assumptions that enter the approximation scheme used to derive the effective equation of motion. Section~\ref{SecConsequencesForWEP} is devoted to the discussion of the consequences of our results in the context of the weak equivalence principle. For the purpose of clarity, in subsection~\ref{SubSecDefinitionOfEP} we first provide the definitions of various flavors of the equivalence principle. Then, in subsection~\ref{SubSecEPinQG} we discuss the status of the equivalence principle in the context of quantum gravity and the results obtained in section~\ref{SecGeodesicEquationInQG}. Subsection~\ref{SubSecUniversalityAndGravitationalMass} provides further analysis of universality and equality between inertial and gravitational masses, in the context of the Newtonian approximation. Finally, section~\ref{SecConclusions} contains our conclusions, discussion of the results and possible lines of further research. In the Appendix
we give a short review of the multipole formalism used in the main text, with some mathematical details.

Our notation and conventions are as follows. We will work in the natural system of units in which $c=\hbar=1$  and $G = l_p^2$, where $l_p$ is the Planck length and $G$ is the Newton's gravitational constant. By convention, the metric of spacetime will have the spacelike Lorentz signature $(-,+,+,+)$. The spacetime indices are denoted with lowercase Greek letters $\mu,\nu,\dots$ and take the values $0,1,2,3$. These can be split into the timelike index $0$ and the spacelike indices denoted with lowercase Latin letters $i,j,k,\dots$ which take the values $1,2,3$. The Lorentz-invariant metric tensor is denoted as $\eta_{\mu\nu}$. Quantum operators always carry a hat, $\hat{\phi}(x)$, $\hat{g}(x)$, etc. The parentheses around indices indicate symmetrization with respect to those indices, while brackets indicate antisymmetrization:
$$
A_{(\mu\nu)} \equiv \frac{1}{2} \left( A_{\mu\nu} + A_{\nu\mu} \right)\,, \qquad
A_{[\mu\nu]} \equiv \frac{1}{2} \left( A_{\mu\nu} - A_{\nu\mu} \right)\,.
$$
Finally, we will systematically denote the values of functions with parentheses, $f(x)$, while functionals will be denoted with brackets, $F[\phi]$.

\section{\label{SecGeodesicEquationInGR}Geodesic equation in general relativity}

In the context of the classical theories of gravity, like GR, the question of deriving the geodesic equation for a particle has initially been studied by Einstein, Infeld and Hoffmann~\cite{EinsteinInfeldHoffmann}, Mathisson~\cite{Mathisson}, Lub\'anski~\cite{Lubanski}, Fock~\cite{Fock}, and others. Slightly later, the question was revisited in the seminal paper by Papapetrou~\cite{Papapetrou}, with generalizations followed by a number of authors~\cite{WeyssenhoffRaabe,Tulczyjew,Taub,Dixon1,Dixon2,Dixon3,Dixon4,Dixon5,YasskinStoeger,ShirafujiNomuraHayashiOne,ShirafujiNomuraHayashiTwo,VasilicVojinovicDirakovaCestica}, developing the so-called {\em multipole formalism}, see the Appendix~\ref{AppMultipoleFormalism}. Recently, the multipole formalism has been reformulated in a manifestly covariant language and extended from pointlike objects to strings, membranes and further to $p$-branes, with general equations of motion studied in Riemann and Riemann-Cartan spaces~\cite{AragoneDeser,VasilicVojinovicJedan,VasilicVojinovicJHEP,VasilicVojinovicDva,VasilicVojinovicTri,VasilicVojinovicCetiri}. Today, the multipole formalism and the resulting classes of effective equations of motion have found applications in a wide range of topics, from string theory~\cite{VasilicVojinovicPet} to cosmology~\cite{Vasilic} to blackbrane dynamics~\cite{JayArmas,JayArmasDva,JayArmasTri} to elasticity and the studies of the shape of red blood cells in biological systems~\cite{JayArmasCetiri}.

In this section we will demonstrate the application of the multipole formalism in its crudest {\em single pole} approximation, and employ it to derive the geodesic equation of motion for a point particle in classical Riemannian spacetime. The results presented in this section are well known in the literature, and illustrate the derivation procedure of the geodesic motion for a point particle. After reviewing the standard results in this section, in section~\ref{SecGeodesicEquationInQG} the same procedure will be utilized to study the quantum gravity case.

The derivation procedure is based on two main assumptions. The first assumption is that the matter fields have internal dynamics such that they form particle-like kink solutions which are stable (i.e., non-decaying) across the spacetime regions under consideration.  If that is the case, one can employ the multipole formalism and expand the stress-energy tensor into a series of derivatives of the Dirac $\delta$ function as (see the Appendix~\ref{AppMultipoleFormalism} for details):
\begin{equation} \label{DeltaExpansionForTEI}
T^{\mu\nu}(x) = \int_{\cC} d\tau \left[ B^{\mu\nu}(\tau) \frac{\delta^{(4)}(x-z(\tau))}{\sqrt{-g}} + \nabla_{\rho} \left( B^{\mu\nu\rho}(\tau) \frac{\delta^{(4)}(x-z(\tau))}{\sqrt{-g}} \right) + \dots \right] \,.
\end{equation}
Here we assume that the stress-energy tensor of matter fields has nonzero value only near some timelike curve $\cC$ represented by parametric equations $x^{\mu}=z^{\mu}(\tau)$, where $\tau$ is a parameter counting the points along the curve $\cC$. In that case, the $B$-coefficients in the $\delta$ series will be smaller and smaller with each new term in the series. We introduce a series of smallness scales for the coefficients,
\begin{equation} 
B^{\mu\nu} \sim \cO_0\,, \qquad B^{\mu\nu\rho} \sim \cO_1\,, \qquad B^{\mu\nu\rho\sigma} \sim \cO_2\,, \qquad \dots \nonumber
\end{equation}
such that one can consider the multipole scales to behave as
\begin{equation} 
\label{eq:orderO}
\cO_0 \gg \cO_1 \gg \cO_2 \gg \dots
\end{equation}
Next we choose to work in the so-called {\em single pole} approximation, in which all quantities of order $\cO_1$ and higher can be neglected. It is also assumed that the typical radius of curvature of spacetime near the curve $\cC$ will be large enough not to interfere in the internal dynamics of the matter fields along $\cC$ and break the kink configuration apart. Physically speaking, the sequence of inequalities~(\ref{eq:orderO}) states that one can systematically approximate the full solution of the matter field equations of motion by neglecting various degrees of freedom which describe the ``size'' and ``shape'' of the kink compared to its orbital motion (i.e., motion along the curve $\cC$). Given this setup, in the single pole approximation the matter fields are in a configuration that looks like a point particle traveling along a worldline curve $\cC$, and terms of order $\cO_1$ and higher can be dropped from the stress-energy tensor, giving:
\begin{equation} 
\label{SinglePoleTEI}
T^{\mu\nu}(x) = \int_{\cC} d\tau \, B^{\mu\nu}(\tau) \frac{\delta^{(4)}(x-z(\tau))}{\sqrt{-g}}\,.
\end{equation}

The second assumption is the validity of the local Poincar\'e invariance for the matter field equations. Namely, the classical action which describes the gravity-matter system can be generally written as
\begin{equation} 
S[g,\phi] = S_G[g] + S_M[g,\phi]\,, \nonumber
\end{equation}
where $g$ and $\phi$ denote gravitational and matter degrees of freedom, respectively, and it is generally considered to feature local Poincar\'e invariance. Our assumption is that the matter action $S_M$ and the gravitational action $S_G$ are invariant even taken separately. If this is the case, the Noether theorem gives us the covariant conservation of the stress-energy tensor of matter fields,
\begin{equation} 
\label{CovariantConservationTEI}
\nabla_{\nu} T^{\mu\nu} = 0\,.
\end{equation}
Taken together, assumptions (\ref{SinglePoleTEI}) and (\ref{CovariantConservationTEI}) are sufficient to establish two results:
\begin{itemize}
\item[(a)] that the parametric functions $z(\tau)$ of the curve $\cC$ satisfy the geodesic equation,
\begin{equation} 
\label{TraditionalGeodesicEquation}
\frac{d^2 z^{\lambda}(\tau)}{d\tau^2} + \mitGamma^{\lambda}{}_{\mu\nu} \frac{d z^{\mu}(\tau)}{d\tau} \frac{d z^{\nu}(\tau)}{d\tau} = 0\,,
\end{equation}
where $\mitGamma^{\lambda}{}_{\mu\nu}$ is the Christoffel connection for the background spacetime metric $g_{\mu\nu}$, and
\item[(b)] that the leading order coefficient $B^{\mu\nu}(\tau)$ in the stress-energy tensor for the particle has the form
\begin{equation} 
\label{FinalStructureOfB}
B^{\mu\nu}(\tau) = m\, u^{\mu}(\tau) u^{\nu}(\tau)\,,
\end{equation}
where $m\in\realni \backslash \{0\}$ is an arbitrary constant parameter, while $u^{\mu}$ is the normalized tangent vector to the curve $\cC$,
$$
u^{\mu} \equiv \frac{dz^{\mu}(\tau)}{d\tau}\,, \qquad u^{\mu} u^{\nu} g_{\mu\nu} = -1\,.
$$
\end{itemize}

In order to demonstrate these two statements, we start from (\ref{CovariantConservationTEI}), contract it with an arbitrary test function $f_{\mu}(x)$ of compact support, and integrate over the whole spacetime,
$$
\int_{\cM_4} d^4x \sqrt{-g}\, f_{\mu} \nabla_{\nu} T^{\mu\nu} = 0\,.
$$
Then we perform the partial integration to move the covariant derivative from the stress-energy tensor to the test function. The boundary term vanishes since the test function has compact support, giving
$$
\int_{\cM_4} d^4x \sqrt{-g}\, T^{\mu\nu} \nabla_{\nu} f_{\mu} = 0\,.
$$
Then we substitute (\ref{SinglePoleTEI}), switch the order of integrations and perform the integral over spacetime $\cM_4$, ending up with
\begin{equation}\label{TransformedCovariantConservationTEI}
\int_{\cC} d\tau B^{\mu\nu} \nabla_{\nu} f_{\mu} = 0 \,.
\end{equation}
The spacetime covariant derivative of the test function can be split into a component tangent to the curve $\cC$ and a component orthogonal to it, in the following way. Using the identity
\begin{equation} 
\label{ProjectorIdentity}
\delta^{\lambda}_{\mu} = - u^{\lambda} u_{\mu} + P_{\orto}^{\lambda}{}_{\mu}\,,
\end{equation}
where $ - u^{\lambda} u_{\mu}$ and $ P_{\orto}^{\lambda}{}_{\mu}$ are projectors along $u^{\mu}$ and orthogonal to $u^{\mu}$, respectively, we rewrite the derivative of $f_{\nu}$ as
\begin{equation} \label{ExpandedTestFunctionDerivative}
\nabla_{\nu} f_{\mu} = - u_{\nu} \nabla f_{\mu} + f_{\nu\mu}^{\orto}\,,
\end{equation}
where $\nabla \equiv u^{\lambda}\nabla_{\lambda}$ is the covariant derivative in the direction of the curve $\cC$, while $f_{\nu\mu}^{\orto} \equiv P_{\orto}^{\lambda}{}_{\nu} \nabla_{\lambda} f_{\mu}$ is a quantity orthogonal to the curve $\cC$ with respect to its first index. Substituting (\ref{ExpandedTestFunctionDerivative}) into (\ref{TransformedCovariantConservationTEI}), and performing another partial integration, we find
$$
\int_{\cC} d\tau \Big[ f_{\mu} \nabla(B^{\mu\nu}u_{\nu}) +  B^{\mu\nu} f_{\nu \mu}^{\orto}\Big] = 0\,,
$$
where the boundary term again vanishes due to the compact support of the test function.

Given that the values of $f_{\mu}$ and $f_{\nu\mu}^{\orto}$ are both arbitrary and mutually independent along the curve $\cC$, the coefficients multiplying them must each be zero. The first term gives us
\begin{equation} 
\label{TermMultiplyingF}
\nabla(B^{\mu\nu}u_{\nu})=0\,,
\end{equation}
while the second term, knowing that $f_{\nu\mu}^{\orto}$ is orthogonal to the curve $\cC$ in its first index, gives
\begin{equation} \label{TermMultiplyingFort}
B^{\mu\nu} P_{\orto}^{\lambda}{}_{\nu} = 0\,.
\end{equation}
Focus first on (\ref{TermMultiplyingFort}). Knowing that $B^{\mu\nu}$ is symmetric, we can use (\ref{ProjectorIdentity}) to decompose it into orthogonal and parallel components with respect to its two indices,
$$
B^{\mu \nu}=B^{\mu \nu}_{\orto}+B^{\mu}_{\orto} u^{\nu} + B^{\nu}_{\orto} u^{\mu} + B u^{\mu} u^{\nu}\,,
$$
where $B^{\mu \nu}_{\orto}$, $B^{\mu}_{\orto}$ and $B$ are unknown coefficients, the first two being orthogonal to the curve $\cC$ in all their indices. Substituting this expansion into (\ref{TermMultiplyingFort}), one finds that
$$
B^{\mu \nu}_{\orto} = 0\,, \qquad B^{\mu}_{\orto} = 0\,,
$$
leaving the scalar $B$ as the only nonzero component of $B^{\mu\nu}$. Changing the notation from $B$ to $m$, one obtains
\begin{equation} \label{NonFinalStructureForB}
B^{\mu \nu} (\tau) = m(\tau) u^{\mu} u^{\nu}\,.
\end{equation}
This equation looks very similar to (\ref{FinalStructureOfB}) but is still not equivalent to it, since the coefficient $m(\tau)$ is still not known to be a constant.

Next, focus on (\ref{TermMultiplyingF}). Substituting (\ref{NonFinalStructureForB}), it reduces to
\begin{equation} \label{AlmostGeodesicEquation}
\nabla (m u^{\mu})= 0\,.
\end{equation}
Projecting onto the tangent direction $u_{\mu}$ and using the identity $u_{\mu} \nabla u^{\mu} = 0$, one obtains
\begin{equation} \label{MassConservation}
\nabla m \equiv \frac{dm}{d\tau} = 0\,,
\end{equation}
establishing that the parameter $m$ is actually a constant. Given this, equation (\ref{AlmostGeodesicEquation}) reduces to
\begin{equation} \label{GeodesicEquation}
\nabla u^{\mu} = 0\,.
\end{equation}
Remembering that $\nabla \equiv u^{\lambda} \nabla_{\lambda}$ and expanding the covariant derivative, we see that this is the geodesic equation (\ref{TraditionalGeodesicEquation}). Finally, (\ref{MassConservation}) and (\ref{NonFinalStructureForB}) together give (\ref{FinalStructureOfB}), which completes the proof of statements (a) and (b).

There are three general remarks one should make regarding the above procedure. The first remark is about the physical interpretation and properties of the free parameter $m$. Namely, it can be given the interpretation of the total mass of the particle --- substituting (\ref{FinalStructureOfB}) into the stress-energy tensor (\ref{SinglePoleTEI}) and integrating the $T^{00}$ component over the volume of the spatial hypersurface orthogonal to $u^{\mu}$, one can easily verify that the total rest-energy of the matter fields at a given time is equal to $m$. Note, however, that the sign of $m$ is not fixed to be positive. This is not surprising, since the covariant conservation equation (\ref{CovariantConservationTEI}) and the stress-energy tensor (\ref{SinglePoleTEI}) do not contain any information (or assumption) about the positivity of energy. Instead, the positive energy condition $m>0$ has to be established from the full matter field equations, which take into account the internal dynamics of the matter fields that make up the particle.


The second remark is about the metric $g_{\mu\nu}$ of the background geometry. When discussing the motion of a particle, the background geometry is usually assumed to be fixed, and backreaction of the gravitational field of the particle itself is not taken into account, leading to the notion of a ``test particle''. However, ignoring the backreaction is not a necessary assumption. Namely, one can take the full stress-energy tensor of the matter fields which form the kink solution (as opposed to the approximate single pole stress-energy tensor (\ref{SinglePoleTEI})), put it as a source into the Einstein's field equations and solve for the metric $g_{\mu\nu}$. The resulting metric does include the backreaction, and can then be reinserted into the geodesic equation for the particle. Note that this procedure is self-consistent, since the geodesic motion of the particle is a consequence of the covariant conservation equation (\ref{CovariantConservationTEI}) which is in turn itself a consequence of Einstein's field equations. Also note that the metric $g_{\mu\nu}$ obtained in this way does not necessarily give rise to the black hole geometry in the neighborhood of the particle. This is because the Schwarzschild radius of the kink may be (and usually is) much smaller than the scale $\cO_1$ which defines the precision of the single pole approximation (\ref{SinglePoleTEI}). A simple example would be the motion of a planet around the Sun --- in the single pole approximation, the radius of the planet (itself far larger than the planet's gravitational radius) is considered to be of the order $\cO_1$ and the planet is treated as a pointlike object, but the spacetime metric used in the geodesic equation can still take into account the planet's gravitational field in addition to the field of the Sun.

The third remark is about going beyond the single pole approximation. This has been studied in detail in the literature~\cite{Papapetrou,WeyssenhoffRaabe,Tulczyjew,Taub,Dixon1,Dixon2,Dixon3,Dixon4,Dixon5,YasskinStoeger,ShirafujiNomuraHayashiOne,ShirafujiNomuraHayashiTwo,VasilicVojinovicJHEP,VasilicVojinovicDva,VasilicVojinovicTri,VasilicVojinovicCetiri}, so here we merely point out the main physical interpretation. Namely, keeping the second term in the multipole expansion (\ref{DeltaExpansionForTEI}) physically amounts to giving the particle a nonzero ``thickness'', in the sense that its internal angular momentum can be considered nonzero. In the resulting equation of motion for the particle, this angular momentum couples to the spacetime curvature tensor, giving rise to a deviation from the geodesic motion. This can intuitively be understood as an effect of tidal forces acting across the scale of the kink's width, pushing it off the geodesic trajectory. Similarly, including quadrupole and higher order terms in  (\ref{DeltaExpansionForTEI}) takes into account additional internal degrees of freedom of the kink, which also couple to spacetime geometry and produce a further deviation from geodesic motion.

\bigskip

The above review of the multipole formalism, and its application to the derivation of the geodesic equation in GR, will be used in the next section to discuss the corrections to the motion of a particle stemming from quantum gravity. As we shall see, these quantum corrections will give rise to additional terms in the effective equation of motion for a particle, pushing it slightly off the geodesic trajectory, even in the single pole approximation.

\section{\label{SecGeodesicEquationInQG}Geodesic equation in quantum gravity}

In this section we discuss the motion of a particle within the framework of quantum gravity. The exposition is structured into four parts --- first, we introduce the abstract quantum gravity formalism, and give some technical details about the description of the states. In the second part, we discuss the quantum version of the covariant conservation equation of the stress-energy tensor. In the third part we adapt the derivation presented in section~\ref{SecGeodesicEquationInGR} to the quantum formalism, and obtain the effective equation of motion for the particle. Finally, in the fourth part we discuss the self-consistency assumptions that go into the calculation.

\subsection{\label{SubSecPrelimiaries}Preliminaries and the setup}

We work in the so-called generic abstract quantum gravity setup, as follows. Starting from the Heisenberg picture for the description of quantum systems, we assume that gravitational degrees of freedom are described by some gravitational field operators $\hat{g}(x)$, while matter degrees of freedom are described by matter field operators $\hat{\phi}(x)$, where $x$ represents the coordinates of some point on a $4$-dimensional spacetime manifold $\cM_4$. Both sets of operators have their corresponding canonically conjugate momentum operators, $\hat{\pi}_g(x)$ and $\hat{\pi}_{\phi}(x)$, such that the usual canonical commutation relations hold. The total (kinematical) Hilbert space of the theory is $\cH_{\rm kin} = \cH_G \otimes \cH_M$, where the gravitational and matter Hilbert spaces $\cH_G$ and $\cH_M$ are spanned by the bases of eigenvectors for the operators $\hat{g}$ and $\hat{\phi}$, respectively. The total state of the system, $\ket{\Psi}\in \cH_{\rm kin}$, does not depend on $x$, in line with the Heisenberg picture framework.

There are several important points that need to be emphasized regarding the above setup. First, we do not explicitly state what are the fundamental degrees of freedom $\hat{g}$ for the gravitational field. They can be chosen in many different ways, giving rise to different models of quantum gravity. Since we aim to present the analysis of geodesic motion which is model-independent, we refrain from specifying what are the fundamental degrees of freedom $\hat{g}$. Instead, we merely assume that the operators describing the spacetime geometry, i.e., the metric, connection, curvature, etc., depend somehow on $\hat{g}$ and $\hat{\pi}_g$, and are expressible as operator functions in terms of them:
$$
\hat{g}_{\mu\nu} = \hat{g}_{\mu\nu} (\hat{g},\hat{\pi}_g)\,, \qquad \hat{\itGamma}^{\lambda}{}_{\mu\nu} = \hat{\itGamma}^{\lambda}{}_{\mu\nu} (\hat{g},\hat{\pi}_g)\,, \qquad \hat{R}^{\lambda}_{\mu\nu\rho} = \hat{R}^{\lambda}_{\mu\nu\rho} (\hat{g},\hat{\pi}_g)\,, \qquad \dots
$$
When discussing these geometric operators, for simplicity we will usually not explicitly write their $(\hat{g},\hat{\pi}_g)$-de\-pen\-den\-ce.

Second, in order for any operator function to be well defined, some operator ordering has to be assumed. However, since we aim to work in an abstract model-independent QG formalism, we do not choose any particular ordering, but merely assume that one such ordering has been fixed. In a similar fashion, we also simply assume that all operators and spaces are well defined, convergent, and otherwise specified in enough mathematical detail to have a well defined and unique QG model. In a nutshell, our calculations are formal, in the sense that one should be able to repeat them in a detailed fashion if one is given a specific model of QG. This also means that our analysis and results should not depend on any of these details, but are rather common to a large class of QG models, and are based only on very few assumptions given above.

Third, we employ a natural distinction between gravitational and matter degrees of freedom. Namely, whereas geometric operators such as metric, curvature, and so on, depend only on $\hat{g}$ and $\hat{\pi}_g$, matter operators like field strengths, stress-energy tensor, etc., will generically be operator functions of both $\hat{g}$, $\hat{\pi}_g$, and the fundamental matter degrees of freedom $\hat{\phi}$ and $\hat{\pi}_{\phi}$. In other words, we assume that the separation between gravity and matter present in the classical theory, described by an action of the form
$$
S_{\rm total}[g,\phi] = S_{\rm gravity}[g] + S_{\rm matter}[g,\phi]\,,
$$
remains present also in the full quantum regime. That is to say, we assume that one can construct a theory of quantum gravity without matter fields, using only gravitational degrees of freedom $g$, so that this theory gives sourceless Einstein's equations of GR in the classical limit. Once such a pure-QG model has been constructed, we assume one can couple matter $\phi$ to it without changing the structure of the gravitational sector, obtaining the full QG model which features Einstein's equations with appropriate matter sources in the classical limit. While we do not consider this to be a strong assumption, we feel that it is nevertheless important to spell it out explicitly, since there may exist some QG models which fail to satisfy it, and our analysis may be inapplicable to such models.

After the introduction of the above conceptual setup, we turn to some more practical details. For the purpose of discussing geodesic motion, we are mostly interested in the effective classical theory of the abstract QG introduced above. To that end, the main objects of attention are {\em classical states} of gravity and matter, denoted by $\ket{\Psi} \in \cH_G \otimes \cH_M$. By classical, we mean that the ``effective classical'' values for the metric tensor and the matter stress-energy tensor, given by the expectation values of the corresponding operators
\begin{equation} \label{gTdef}
g_{\mu\nu} = \bra{\Psi} \hat{g}_{\mu\nu} \ket{\Psi}\,, \qquad T_{\mu\nu} =\bra{\Psi} \hat{T}_{\mu\nu} \ket{\Psi}\,
\end{equation}
satisfy classical Einstein equations of the GR. A recent study suggests that physical states of gravity and matter are generically entangled~\cite{PaunkovicVojinovic2017}. For our analysis, we do not need to assume that the overall gravity-matter state is separable, and thus we will work with a generic state $\ket{\Psi}$ (see Appendix~\ref{AppProductCoherent} for the discussion of the separable case).

For the purpose of our paper, we will consider a \textit{toy example state}, defined as
\begin{equation} \label{eq:ent_st_1}
\ket{\PsiEnt} = \alpha \ket{\Psi} + \beta \ket{\tilde{\Psi}}\,,
\end{equation}
where $\ket{\tilde{\Psi}}$ is some other classical state analogous to $\ket{\Psi}$, but giving different expectation values for the classical metric and stress-energy tensors:
\begin{equation} \label{gTdefTilde}
\tilde{g}_{\mu\nu} = \bra{\tilde{\Psi}} \hat{g}_{\mu\nu} \ket{\tilde{\Psi}}\,, \qquad
\tilde{T}_{\mu\nu} = \bra{\tilde{\Psi}} \hat{T}_{\mu\nu} \ket{\tilde{\Psi}}\,.
\end{equation}
One can see that our toy-example state (\ref{eq:ent_st_1}) is a Schr\"odinger-cat type of state, describing a coherent superposition of two classical configurations of gravitational and matter fields. It will become evident later on that qualitative conclusions of the paper do not depend on the fact that (\ref{eq:ent_st_1}) has precisely two terms in the sum. Choosing the state with three, four or more terms will lead to analogous conclusions, although quantitative details of the computation may become technically more involved.

Given that (\ref{eq:ent_st_1}) is a Schr\"odinger-cat type of state, there are some phenomenological restrictions on the values of the independent parameters $\beta$ and $S \equiv \bracket{\Psi}{\tilde{\Psi}}$. Namely, in the ordinary experimental situations we basically never observe this kind of states, which means that the overall entangled state $\ket\PsiEnt$ looks pretty much like a classical state, say the state $\ket{\Psi}$. In other words, we want the fidelity between these two states to be large, $F(\ket\Psi,\ket\PsiEnt) = |\bracket{\Psi}{\PsiEnt}| \approx 1$. From (\ref{eq:ent_st_1}) we obtain
\begin{equation}
\bracket{\Psi}{\PsiEnt} = \alpha + \beta S \equiv \kappa\,. \nonumber
\end{equation}
Define $\ket{\Psi^{\orto}}$, such that 
\begin{equation}
\ket{\tilde{\Psi}} = S \ket{\Psi} + \epsilon \ket{\Psi^{\orto}}\,, \nonumber
\end{equation}
$\bracket{\Psi}{\Psi^{\orto}} = 0$ and $\epsilon = \sqrt{1 - |S|^2}$. Thus,
\begin{equation} \label{FidelityRequirement}
F^2 = |\bracket{\Psi}{\PsiEnt}|^2 = 1 - \eta^2\,,
\end{equation}
where we introduce the small parameter
\begin{equation}
\eta \equiv \beta\epsilon\,. \nonumber
\end{equation}
Here, we have used the normalization condition for the entangled state (\ref{eq:ent_st_1}),
\begin{equation}
\bracket{\PsiEnt}{\PsiEnt} = \alpha^2 + \beta^2 + 2\alpha\beta \Re(S) = 1\,. \nonumber
\end{equation}
Given the above definitions for $\kappa$ and $\eta$, we can rewrite the total state (\ref{eq:ent_st_1}) as
\begin{equation} \label{FullEntangledState}
\ket{\PsiEnt} = \kappa \ket{\Psi} + \eta \ket{\Psi^{\orto}} \,,
\end{equation}
where $|\kappa|^2 = F^2 = 1-\eta^2$. The physical requirement of large fidelity implies that we study the limit $|\kappa| \approx 1$, $\eta\to 0$. We will therefore systematically expand the expectation values of all operators into power series in $\eta$, up to order $\cO(\eta^2)$. 

At this point we can evaluate the expectation values for the metric and stress-energy operators in the state (\ref{FullEntangledState}),  obtaining
\begin{equation} \label{MetricExpansionInBetaExact}
\gEnt_{\mu\nu} = \bra{\PsiEnt}\hat{g}_{\mu\nu}\ket{\PsiEnt} = ( 1-\eta^2) g_{\mu\nu} + \eta^2 \bra{\Psi^{\orto}} \hat{g}_{\mu\nu} \ket{\Psi^{\orto}} + 2 \eta \Re \left( \kappa \bra{\Psi^{\orto}} \hat{g}_{\mu\nu} \ket{\Psi} \vphantom{\hat{T}} \right)\,,
\end{equation}
\begin{equation} \label{TEIexpansionInBetaExact}
\TEnt_{\mu\nu} = \bra{\PsiEnt}\hat{T}_{\mu\nu}\ket{\PsiEnt} = (1-\eta^2) T_{\mu\nu} + \eta^2 \bra{\Psi^{\orto}} \hat{T}_{\mu\nu} \ket{\Psi^{\orto}} + 2\eta \Re \left( \kappa \bra{\Psi^{\orto}} \hat{T}_{\mu\nu} \ket{\Psi}\right) \,.
\end{equation}
It is easy to see that interference terms from the above expressions are generically nonvanishing. Indeed, even if, say, $\kappa \bra{\Psi^{\orto}} \hat{g}_{\mu\nu} \ket{\Psi} \vphantom{\hat{T}}$ were purely imaginary, a simple change of relative phase between $\ket{\Psi}$ and $\ket{\tilde{\Psi}}$ would give rise to a nontrivial real part. Namely, given a fixed choice of $\ket{\tilde{\Psi}}$, the set of choices for $\ket{\Psi}$ for which the interference term is purely imaginary is of measure zero compared to the full set of possible phase shifts of $\ket{\Psi}$. An analogous argument applies for $\kappa \bra{\Psi^{\orto}} \hat{T}_{\mu\nu} \ket{\Psi}$ as well. For a detailed analysis, see Appendix~\ref{AppDiscussionOfTheAngle}.

Let us denote the metric and stress-energy interference terms as $\bar{g}_{\mu\nu}$ and $\bar{T}_{\mu\nu}$, respectively. Since we want to expand (\ref{MetricExpansionInBetaExact}) and (\ref{TEIexpansionInBetaExact}) into power series in $\eta$ up to linear order, we can write
\begin{equation} \label{MetricOverlap}
\bar{g}_{\mu\nu} \equiv 2 \Re \left( \kappa \bra{\Psi^{\orto}} \hat{g}_{\mu\nu} \ket{\Psi} \vphantom{\hat{T}} \right) = \barg_{\mu\nu} + \cO(\eta) \,,
\end{equation}
\begin{equation} \label{StressEnergyOverlap}
\bar{T}_{\mu\nu} \equiv 2 \Re \left( \kappa \bra{\Psi^{\orto}} \hat{T}_{\mu\nu} \ket{\Psi} \right) = \barT_{\mu\nu} + \cO(\eta)\,.
\end{equation}
Here, $\barg_{\mu\nu}$ and $\barT_{\mu\nu}$ are $\eta$-independent parts of $\bar{g}_{\mu\nu}$ and $\bar{T}_{\mu\nu}$. Thus, we can finally write:
\begin{equation} \label{MetricExpansionInBeta}
\gEnt_{\mu\nu} = g_{\mu\nu} + \eta \, \barg_{\mu\nu} + \cO(\eta^2) \,,
\end{equation}
\begin{equation} \label{TEIexpansionInBeta}
\TEnt_{\mu\nu} = T_{\mu\nu} + \eta \, \barT_{\mu\nu} + \cO(\eta^2)\,.
\end{equation}

In what follows, we will refer to the classical state $\ket{\Psi}$ as the \textit{dominant state}, while the other classical state $\ket{\tilde{\Psi}}$ will be called the \textit{sub-dominant state}. To justify this terminology, recall the above requirement (\ref{FidelityRequirement}) that the overall entangled state $\ket{\PsiEnt}$ looks like the classical state $\ket{\Psi}$, i.e., $F^2 = 1 -\eta^2$, with the parameter $\eta\equiv \beta\epsilon$ being small. Therefore, in the case $\beta\to 0$ and $\epsilon$ finite, the state $\ket{\tilde{\Psi}}$ enters (\ref{eq:ent_st_1}) with a very small contribution, and is thus sub-dominant. On the other hand, in the case when $\beta$ is finite and $\epsilon\to 0$, the states $\ket{\Psi}$ and $\ket{\tilde{\Psi}}$ are essentially indistinguishable, and their roles can be exchanged, as either can be considered sub-dominant to the other. By convention, we choose $\ket{\tilde{\Psi}}$ to again play the role of the sub-dominant state.

While in any quantum theory entangled states are allowed, note that when considering a product state of the gravity-matter system (i.e., the case $\eta=0$), there is a danger that such a state may fail to be gauge invariant, as argued in \cite{PaunkovicVojinovic2017}. So we need to introduce at least a small sub-dominant state, in order to ensure the gauge invariance of the total state. The simplest possible candidate state which describes the classical physics sufficiently well, and simultaneously stands a chance of being gauge invariant, is the genuinely entangled state (\ref{eq:ent_st_1}), with $\beta\neq 0$ and $\ket{\tilde{\Psi}} \neq \ket{\Psi}$, leading to $\eta$ being very small, but nonzero.

Regarding the effective entangled metric and stress-energy tensors (\ref{MetricExpansionInBeta}) and (\ref{TEIexpansionInBeta}), it is important to stress that they do not satisfy classical Einstein's equations of GR. Namely, we assume that Einstein's equations are separately satisfied by the metric and stress-energy tensors (\ref{gTdef}) coming from the classical state $\ket{\Psi}$, and by the metric and stress-energy tensors (\ref{gTdefTilde}) coming from the other classical state $\ket{\tilde{\Psi}}$, as two different classical solutions:
$$
R_{\mu\nu}(g) - \frac{1}{2}g_{\mu\nu} R(g) = 8\pi l_p^2 \, T_{\mu\nu}\,, \qquad
R_{\mu\nu}(\tilde{g}) - \frac{1}{2}\tilde{g}_{\mu\nu} R(\tilde{g}) = 8\pi l_p^2 \, \tilde{T}_{\mu\nu}\,.
$$
However, due to the nonlinearity of Einstein's equations, and due to the presence of the interference terms $\barg_{\mu\nu}$ and $\barT_{\mu\nu}$ in (\ref{MetricExpansionInBeta}) and (\ref{TEIexpansionInBeta}), quantities $\gEnt_{\mu\nu}$ and $\TEnt_{\mu\nu}$ do not satisfy Einstein's equations, as long as $\eta\neq 0$. This leads us to the following physical interpretation. First, it is natural to expand all quantities as corrections to the dominant classical configuration  $(g_{\mu\nu},T_{\mu\nu})$, including the equation of motion for a point particle. Second, as we shall see in the remainder of the text, given that $(\gEnt_{\mu\nu},\TEnt_{\mu\nu})$ contains quantum gravity corrections through the interference terms, the presence of these quantum corrections in (\ref{MetricExpansionInBeta}) and (\ref{TEIexpansionInBeta}) will introduce an ``effective force'' term into the effective equation of motion for the particle. Finally, this effective force term will be pushing the particle off the geodesic trajectory defined by the classical dominant metric $g_{\mu\nu}$.

\subsection{\label{SubSecCovariantConservationEquation}Effective covariant conservation equation}

After the discussion of the general QG setup and the state (\ref{eq:ent_st_1}), we move on to the discussion of the quantum analog of the covariant conservation equation (\ref{CovariantConservationTEI}). As in the classical theory, our basic assumption is that the matter sector of our QG model features local Poincar\'e invariance, i.e., that this symmetry is preserved at the quantum level. This assumption gives rise to a Gupta-Bleuler-like condition on the physical states, in the form
\begin{equation} \label{QuantumCovariantConservationTEI}
\bra{\PsiEnt} \hat{\nabla}_{\nu} \hat{T}^{\mu\nu} \ket{\PsiEnt} = 0\,,
\end{equation}
where $\hat{\nabla}_{\mu}$ is the covariant derivative operator, defined by promoting the metric appearing in the Christoffel symbols into a corresponding operator. In general, the action of the stress-energy operator on the state $\ket{\PsiEnt}$ can be written\footnote{Given any self-adjoint operator $\hat{A}$ and any state $\ket{\Psi}$, one can write
$$
\hat{A} \ket{\Psi} = a \ket{\Psi} + b \ket{\Psi^{\orto}}\,,
$$
where $\bracket{\Psi}{\Psi^{\orto}}\equiv 0$ and $a,b\in\kompleksni$. Multiplying this equation by $\bra{\Psi}$ and by $\bra{\Psi}\hat{A}$ from the left, one easily obtains that $a$ and $b$ are the expectation value and the uncertainty of the operator $\hat{A}$ in the state $\ket{\Psi}$, respectively.} as
\begin{equation} \label{StressEnergyOperatorActingOnTheState}
\hat{T}^{\mu\nu} \ket{\PsiEnt} = \TEnt^{\mu\nu} \ket{\PsiEnt} + \DeltaTEnt^{\mu\nu} \ket{\PsiEnt^{\orto}}\,,
\end{equation}
where $\TEnt^{\mu\nu}$ and $\DeltaTEnt^{\mu\nu}$ are the expectation value and the uncertainty of the operator $\hat{T}^{\mu\nu}$ in the state $\ket{\PsiEnt}$, respectively,
$$
\TEnt^{\mu\nu} \equiv \bra{\PsiEnt} \hat{T}^{\mu\nu} \ket{\PsiEnt}\,, \qquad
\DeltaTEnt^{\mu\nu} \equiv \sqrt{ \bra{\PsiEnt} \left( \hat{T}^{\mu\nu} \right)^2 \ket{\PsiEnt} - \left( \bra{\PsiEnt} \hat{T}^{\mu\nu} \ket{\PsiEnt} \right)^2 }\,,
$$
while $\ket{\PsiEnt^{\orto}}$ is some state orthogonal to $\ket{\PsiEnt}$. Note that the equation of the form (\ref{StressEnergyOperatorActingOnTheState}) is completely general, holding for any stress-energy operator acting on an arbitrary state. Substituting (\ref{StressEnergyOperatorActingOnTheState}) into (\ref{QuantumCovariantConservationTEI}), we obtain
\begin{equation} \label{QuantumCovariantConservationTEIrewritten}
\nablaEnt_{\nu} \TEnt^{\mu\nu} + \bra{\PsiEnt} \hat{\nabla}_{\nu}  \ket{\PsiEnt^{\orto}} \DeltaTEnt^{\mu\nu} = 0\,,
\end{equation}
where $\nablaEnt_{\nu}$ is the expectation value of the operator $\hat{\nabla}_{\nu}$,
$$
\nablaEnt_{\nu} \equiv \bra{\PsiEnt} \hat{\nabla}_{\nu} \ket{\PsiEnt}\,.
$$

At this point we need to make one more assumption. Namely, we assume that the error scale of the single pole approximation is bigger than the uncertainty of the stress-energy operator, $\DeltaTEnt^{\mu\nu}$. Symbolically,
\begin{equation} \label{RequirementOnStressEnergyUncertainty}
\cO_1 \gtrsim \DeltaTEnt^{\mu\nu} \,.
\end{equation}
This means that in the single pole approximation we do not see the effects of the quantum fluctuations of matter fields. Intuitively, this is a reasonable assumption in most cases. For example, in the case of the kink solution describing the hydrogen atom, the scale on which one can detect quantum fluctuations (i.e., the Lamb shift effects) is much smaller than the size of the atom itself (i.e., the radius of the first Bohr orbit). Therefore, we expect that if our single pole approximation ignores the size of the atom itself, it also ignores the corresponding quantum fluctuations. An analogous assumption is made in relation to the uncertainty of the metric operator $\hat{g}_{\mu\nu}$,
\begin{equation} \label{RequirementOnMetricUncertainty}
\cO_1 \gtrsim \DeltagEnt_{\mu\nu} \,,
\end{equation}
given that the quantum gravity fluctuations can arguably also be ignored in the single pole approximation.

Applying (\ref{RequirementOnStressEnergyUncertainty}) to (\ref{QuantumCovariantConservationTEIrewritten}), in the single pole approximation the second term can be dropped, leading to the effective classical covariant conservation equation,
\begin{equation} \label{EffectiveClassicalEntangledCovariantConservationTEI}
\nablaEnt_{\nu} \TEnt^{\mu\nu} = 0\,.
\end{equation}
In a similar fashion, one can employ (\ref{RequirementOnMetricUncertainty}) to drop the off-diagonal components in the Christoffel symbol operators, leading to an effective classical expression
\begin{equation} \label{EffectiveClassicalEntangledChristoffel}
\GammaEnt{\lambda}{\mu\nu} = \frac{1}{2} \gEnt^{\lambda\sigma} \left( \del_{\mu} \gEnt_{\sigma\nu} + \del_{\nu} \gEnt_{\sigma\mu} - \del_{\sigma} \gEnt_{\mu\nu}  \right)\,,
\end{equation}
where $\gEnt_{\mu\nu} \equiv \bra{\PsiEnt} \hat{g}_{\mu\nu} \ket{\PsiEnt}$ is the effective classical metric and $\gEnt^{\mu\nu}$ is its inverse matrix.

With effective classical expressions (\ref{EffectiveClassicalEntangledCovariantConservationTEI}) and (\ref{EffectiveClassicalEntangledChristoffel}) in hand, we can now employ (\ref{MetricExpansionInBeta}) and (\ref{TEIexpansionInBeta}) to expand them into the dominant and correction parts. First we use (\ref{MetricExpansionInBeta}) and $\gEnt_{\mu\lambda}\gEnt^{\lambda\nu} = \delta^{\nu}_{\mu}$ to find the inverse entangled metric $\gEnt^{\mu\nu} = g^{\mu\nu} - \eta g^{\mu\rho} g^{\nu\sigma} \barg_{\rho\sigma} + \cO(\eta^2)$, and then substitute into (\ref{EffectiveClassicalEntangledChristoffel}) to obtain
\begin{equation} \label{ChristoffelExpansionInBeta}
\GammaEnt{\lambda}{\mu\nu} = \mitGamma^{\lambda}{}_{\mu\nu} + \frac{\eta}{2} g^{\lambda\sigma} \left( \nabla_{\mu} \barg_{\sigma\nu} + \nabla_{\nu} \barg_{\sigma\mu} - \nabla_{\sigma} \barg_{\mu\nu} \right) + \cO(\eta^2)\,,
\end{equation}
where the Christoffel symbols in ordinary $\nabla_{\mu}$ are defined with respect to the dominant classical metric $g_{\mu\nu}$\,. Then, expanding (\ref{EffectiveClassicalEntangledCovariantConservationTEI}) into the form
$$
\del_\nu\TEnt^{\mu\nu}+\GammaEnt{\mu}{\sigma\nu} \TEnt^{\sigma\nu}+\GammaEnt{\nu}{\sigma\nu} \TEnt^{\mu\sigma}=0\,,
$$
we substitute (\ref{TEIexpansionInBeta}) and (\ref{ChristoffelExpansionInBeta}), and after a bit of algebra we rewrite it as:
\begin{equation} \label{EMcorrection2old}
\nabla_{\nu} T^{\mu\nu}+\eta \left[ \nabla_{\nu} \barT^{\mu\nu}+T^{\sigma \nu} \left( \nabla_{\sigma}\barg^{\mu}{}_{\nu}-\frac{1}{2} \nabla^{\mu} \barg_{\nu \sigma} \right) + \frac{1}{2} T^{\mu \sigma} \nabla_{\sigma} h^{\nu}{}_{\nu} \right] + \cO(\eta^2) = 0\,.
\end{equation}
This equation is the one we sought out --- it represents the analog of the classical covariant conservation equation (\ref{CovariantConservationTEI}), while taking into account the interference terms between the two classical states in (\ref{eq:ent_st_1}), approximated to the linear order in $\eta$.

As a final step, (\ref{EMcorrection2old}) can be rewritten in a more compact form. For convenience, introduce the following shorthand notation (see our conventions from the last paragraph of the Introduction),
\begin{equation} \label{defF}
F^{\mu}{}_{\nu\sigma} \equiv \nabla_{(\sigma} \barg^{\mu}{}_{\nu)} -\frac{1}{2} \nabla^{\mu} \barg_{\nu\sigma}\,,
\end{equation}
and also note that
\begin{equation} 
F^{\nu}{}_{\nu\sigma} = \frac{1}{2}\nabla_{\sigma} \barg^{\nu}{}_{\nu} + \frac{1}{2}\nabla_{\nu} \barg^{\nu}{}_{\sigma} -\frac{1}{2} \nabla^{\nu} \barg_{\nu\sigma} = \frac{1}{2}\nabla_{\sigma} \barg^{\nu}{}_{\nu}\,, \nonumber
\end{equation}
so that, dropping the term $\cO(\eta^2)$, equation (\ref{EMcorrection2old}) is rewritten as:
\begin{equation} \label{EMcorrection2}
\nabla_{\nu} \left( T^{\mu\nu} + \eta \barT^{\mu\nu} \right)
 + 2 \eta F^{(\mu}{}_{\nu\sigma} T^{\nu) \sigma} = 0\,.
\end{equation}

The equation (\ref{EMcorrection2}) represents the effective classical covariant conservation law for the stress-energy tensor, with the included quantum correction, represented to first order in $\eta$. It is the starting point for the remainder of our analysis, and replaces equation (\ref{CovariantConservationTEI}) in the derivation of the equation of motion for a point particle.

Finally, note that the quantum correction term in (\ref{EMcorrection2}) has two distinct parts --- one part comes from the quantum correction to the dominant classical stress-energy tensor, i.e., the interference term $\barT^{\mu\nu}$, while the other part comes from the quantum correction to the dominant classical metric, i.e., the interference term $\barg_{\mu\nu}$. This latter quantum correction enters through the Christoffel connection terms present in the covariant derivative. As we shall see in the next subsection, its presence will be crucial for the ``force term'' in the equation of motion for the particle, responsible for the deviation from the classical geodesic trajectory.

\subsection{\label{SubSecEquationOfMotion}Effective equation of motion}

We are now ready to derive the equation of motion for a particle in the single pole approximation, using the technique presented in section~\ref{SecGeodesicEquationInGR}. However, instead of (\ref{CovariantConservationTEI}), we start from the effective covariant conservation law (\ref{EMcorrection2}), which contains the quantum correction terms. Throughout, we assume the following relation of scales,
$$
\cO(\eta) > \cO_1 \geq \cO(\eta^2)\,.
$$
In other words, we assume that the quantum correction terms linear in $\eta$ are not smaller than the width of our particle, since otherwise one could simply ignore them and recover the classical geodesic motion for the particle.

Repeating the method of section~\ref{SecGeodesicEquationInGR}, we begin by contracting (\ref{EMcorrection2}) with an arbitrary test function $f_{\mu}(x)$ of compact support, and integrating over the whole spacetime,
$$
\int_{\cM_4} d^4x \sqrt{-g}\, f_{\mu} \Big[ \nabla_{\nu} \left( T^{\mu\nu} + \eta \barT^{\mu\nu} \right)  + 2 \eta F^{(\mu}{}_{\nu\sigma} T^{\nu) \sigma} \Big] = 0\,.
$$
We then perform the partial integration to move the covariant derivative from the stress-energy tensors to the test function. As before, the boundary term vanishes since the test function has compact support, giving
\begin{equation} \label{PreparedEffectiveCovariantConservationOfTEI}
\int_{\cM_4} d^4x \sqrt{-g}\,  \Big[ - \left( T^{\mu\nu} + \eta \barT^{\mu\nu} \right) \nabla_{\nu} f_{\mu}   + 2  \eta F^{(\mu}{}_{\nu\sigma} T^{\nu) \sigma} f_{\mu} \Big] = 0\,.
\end{equation}

Now we need to model the dominant and correction parts of the stress-energy tensor. For the dominant part, it is straightforward to assume the single pole approximation, as was done in the classical case:
\begin{equation} \label{DeltaExpansionForT}
T^{\mu\nu}(x) = \int_{\cC} d\tau B^{\mu\nu}(\tau) \frac{\delta^{(4)}(x-z(\tau))}{\sqrt{-g}} \,.
\end{equation}
Regarding the correction term, we also use the single pole approximation,
\begin{equation} \label{DeltaExpansionForBarT}
\barT^{\mu\nu}(x) = \int_{\cC} d\tau \bar{B}^{\mu\nu}(\tau) \frac{\delta^{(4)}(x-z(\tau))}{\sqrt{-g}} \,,
\end{equation}
but one should note that in the case of $\barT^{\mu\nu}$ it is less obvious why this approximation is adequate, and requires some justification. However, in order to focus on the derivation of the particle equation of motion, for the moment we simply adopt (\ref{DeltaExpansionForBarT}), and postpone the analysis and the meaning of this approximation for subsection~\ref{SubSecConsistency}.

Then we substitute (\ref{DeltaExpansionForT}) and (\ref{DeltaExpansionForBarT}) into (\ref{PreparedEffectiveCovariantConservationOfTEI}), switch the order of integrations and perform the integral over spacetime $\cM_4$, ending up with
\begin{equation} \label{TransformedCovariantConservationTEIsecondCase}
\int_{\cC} d\tau \Big[ - \left( B^{\mu\nu} + \eta \bar{B}^{\mu\nu} \right) \nabla_{\nu} f_{\mu}   + 2  \eta F^{(\mu}{}_{\nu\sigma} B^{\nu) \sigma} f_{\mu} \Big] = 0\,.
\end{equation}
The next step is to employ the identity (\ref{ExpandedTestFunctionDerivative}) to separate the tangential and orthogonal components of the derivative of the test function. Substituting it into (\ref{TransformedCovariantConservationTEIsecondCase}), and performing another partial integration, we find
$$
\int_{\cC} d\tau \Big[ \left( B^{\mu\nu} + \eta \bar{B}^{\mu\nu} \right) f^{\orto}_{\nu\mu} +  \left[ \nabla \left( B^{\mu\nu} u_{\nu} + \eta \bar{B}^{\mu\nu} u_{\nu} \right) - 2  \eta F^{(\mu}{}_{\nu\sigma} B^{\nu) \sigma} \right] f_{\mu} \Big] = 0\,,
$$
where the boundary term again vanishes due to the compact support of the test function.

After these transformations, we make use of the same argument that both $f_{\mu}$ and $f_{\nu\mu}^{\orto}$ are arbitrary and mutually independent along the curve $\cC$, concluding that the coefficients multiplying them must each be zero. The first term gives us
\begin{equation} \label{TermMultiplyingFsecondCase}
\nabla \left( B^{\mu\nu} u_{\nu} + \eta \bar{B}^{\mu\nu} u_{\nu} \right) - 2  \eta F^{(\mu}{}_{\nu\sigma} B^{\nu) \sigma} = 0\,,
\end{equation}
while the second term, knowing that $f_{\nu\mu}^{\orto}$ is orthogonal to the curve $\cC$ in its first index, gives
\begin{equation} 
\left( B^{\mu\nu} + \eta \bar{B}^{\mu\nu} \right) P_{\orto}^{\lambda}{}_{\nu} = 0\,. \nonumber
\end{equation}
As in the previous case, given that $B^{\mu\nu}$ and $\bar{B}^{\mu\nu}$ are symmetric, one can decompose them into tangential and orthogonal components using (\ref{ProjectorIdentity}), and then from (\ref{TermMultiplyingFort}) read off that all orthogonal components must be zero, concluding that
\begin{equation} \label{NonFinalStructureForBandBarB}
B^{\mu \nu} + \eta \bar{B}^{\mu \nu}  = (B+\eta\bar{B})  u^{\mu} u^{\nu} \equiv m(\tau) u^{\mu} u^{\nu}\,,
\end{equation}
where again we emphasized that the parameter $m$ may depend on the particle's proper time $\tau$.

Next, substituting this into (\ref{TermMultiplyingFsecondCase}) and neglecting the term $\cO(\eta^2)$, we obtain
\begin{equation} \label{AlmostGeodesicEquationWithCorrection}
\nabla \left( m u^{\mu} \right) + \eta m u^{\sigma} \left( F^{\mu}{}_{\nu\sigma} u^{\nu} + F^{\nu}{}_{\nu\sigma} u^{\mu} \right) = 0\,.
\end{equation}
Projecting onto the tangent direction $u_{\mu}$ and using the identity $u_{\mu} \nabla u^{\mu} = 0$, one obtains
\begin{equation} \label{MassNonConservation}
\nabla m = \eta m u^{\sigma} \left( u^{\nu} u_{\lambda} F^{\lambda}{}_{\nu\sigma} - F^{\nu}{}_{\nu\sigma}\right)\,,
\end{equation}
establishing that, in contrast to the classical case, here the parameter $m$ fails to be constant. Substituting this back into the equation (\ref{AlmostGeodesicEquationWithCorrection}), after some simple algebra we obtain
$$
\nabla u^{\mu} + \eta u^{\nu} u^{\sigma} P_{\orto}^{\mu}{}_{\lambda} F^{\lambda}{}_{\nu\sigma} = 0\,,
$$
where the parameter $m$ cancels out of the equation. As a final step, introducing the shorthand notation $F_{\orto}^{\mu}{}_{\nu\sigma} \equiv P_{\orto}^{\mu}{}_{\lambda}F^{\lambda}{}_{\nu\sigma}$, we can rewrite the equation of motion in its final form
\begin{equation} \label{worldlineeq5}
\nabla u^{\mu}+\eta u^{\nu}u^{\sigma}  F_{\orto}^{\mu}{}_{\nu\sigma}=0\,.
\end{equation}
The presence of the orthogonal projector in the second term should not be surprising. Namely, since acceleration must always be orthogonal to the velocity, the second term in the equation must also be orthogonal to velocity, and this is guaranteed by the presence of the orthogonal projector.

Equations (\ref{NonFinalStructureForBandBarB}), (\ref{MassNonConservation}) and (\ref{worldlineeq5}) are the main result of this paper, and we discuss them in turn. Equation (\ref{NonFinalStructureForBandBarB}) determines the structure of the stress-energy tensor describing the point particle, as a function of tangent vectors of its world line and a scalar parameter $m(\tau)$. Formally, it has the same form as its classical counterpart (\ref{NonFinalStructureForB}), and provisionally the parameter $m$ may be even called \textit{effective mass}. Namely, in the rest frame of the particle, integration of the $\TEnt^{00}$ component of the entangled stress-energy tensor over the $3$-dimensional spatial hypersurface can be interpreted as the total rest-energy of the kink configuration of fields that represents the particle. This terminology is of course provisional, since all these notions are merely a part of the semiclassical approximation of the full quantum gravity description.

Equation (\ref{MassNonConservation}) determines the proper time evolution of the parameter $m(\tau)$. In contrast to the classical case, where $m(\tau)$ was determined by (\ref{MassConservation}) to be a constant, here we see that its time derivative is proportional to (covariant derivatives of) the interference term $\barg_{\mu\nu}$ between the dominant and sub-dominant classical geometry, via (\ref{defF}). If one puts $\eta=0$, (\ref{MassNonConservation}) reduces to the classical case, as expected. The interference between the two geometries gives rise to an effective force that is responsible for the change in time of the particle's effective mass. Since the particle is (effectively) not isolated, its total energy is therefore not conserved, in the sense of equation (\ref{MassNonConservation}).

Finally, and most importantly, equation (\ref{worldlineeq5}) represents the effective equation of motion of the particle, determining its world line. It has the form of the classical geodesic equation (\ref{GeodesicEquation}) with an additional correction term proportional to $\eta$ and to the interference term $\barg_{\mu\nu}$. This additional term represents an \textit{effective force}, pushing the particle off the classical geodesic trajectory. It is analogous to the notion of the ``exchange interaction'' force in molecular physics, in the region where the wavefunctions of the two electrons in a molecule overlap.

In our case, however, the force term is determined by the interference between the two classical spacetime and matter configurations superposed in the state (\ref{eq:ent_st_1}), and in particular by the off-diagonal components of the metric operator $\hat{g}_{\mu\nu}$, see (\ref{MetricOverlap}). It is thus a \textit{pure quantum gravity effect}, a consequence of the nontrivial structure of the metric operator. Of course, the detailed properties and the magnitude of the force term depend on the choice of the two classical gravity-matter configurations and on the details of the quantization of the gravitational field.

\subsection{\label{SubSecConsistency}Consistency of the approximation scheme}

Regarding the analysis and the derivation of the effective equation of motion for a particle discussed in the previous subsection, there is one issue that we should reflect on. It is related to the additional consistency conditions that stem from our assumption that the quantum correction to the stress-energy tensor is approximated with a single pole term (\ref{DeltaExpansionForBarT}).

Namely, the two stress-energy tensors that enter the derivation of the effective equation of motion --- the classical dominant stress-energy tensor $T^{\mu\nu}$, and the interference stress-energy tensor $\barT^{\mu\nu}$ --- can in general be written in the single pole approximation as:
\begin{equation} \label{DefinitionOfTheTscale}
T^{\mu\nu} = \int_{\cC} d\tau \, B^{\mu\nu}(\tau) \frac{\delta^{(4)}(x-z(\tau))}{\sqrt{-g}} + \cO_1(T)\,,
\end{equation}
\begin{equation} \label{DefinitionOfTheTbarScale}
\barT^{\mu\nu} = \int_{\cC} d\tau \, \bar{B}^{\mu\nu}(\tau) \frac{\delta^{(4)}(x-z(\tau))}{\sqrt{-g}} + \cO_1(\barT)\,. 
\end{equation}
Note that we have introduced two different $\cO_1$ scales, one for each tensor. This is because, although we assume that both can be expanded into the $\delta$ series around the same curve $\cC$, each tensor may have different ``width'', or in other words, the two configurations of matter fields may be such that they can be well approximated with a single pole term up to a priori two different $\cO_1$ scales. In particular, if one chooses the $\cO_1$ scale to write $T^{\mu\nu}$ in a single pole approximation, $\cO_1 = \cO_1(T)$, it is not obvious that $\barT^{\mu\nu}$ can also be approximated by a single pole term, compared to the same scale, and vice versa. Therefore, there is an assumption about the relationship between scales that we have made when we used expressions (\ref{DeltaExpansionForT}) and (\ref{DeltaExpansionForBarT}) in the derivation of the effective equation of motion.

Looking at the structure of the equation (\ref{PreparedEffectiveCovariantConservationOfTEI}) into which (\ref{DeltaExpansionForT}) and (\ref{DeltaExpansionForBarT}) have been substituted, the consistency condition for the approximation scheme can be written as
\begin{equation} \label{ConsistencyConditionOfScales}
\cO_1 \equiv \cO_1(T) \geq \eta \cO_1(\barT)\,.
\end{equation}
In particular, if this inequality were not valid, the dipole term in (\ref{DefinitionOfTheTbarScale}) would contribute to (\ref{PreparedEffectiveCovariantConservationOfTEI}) with a magnitude comparable to the single pole term of (\ref{DefinitionOfTheTscale}), and it would be inconsistent to ignore it in the derivation of the effective equation of motion.

The consistency condition (\ref{ConsistencyConditionOfScales}) can be rewritten into a more explicit form. Substituting (\ref{DefinitionOfTheTbarScale}) and (\ref{StressEnergyOverlap}) into (\ref{ConsistencyConditionOfScales}), we get
$$
\cO_1(T)\geq \eta \left[ 2 \Re \left( \bra{\Psi^{\orto}} \hat{T}_{\mu\nu} \ket{\Psi}\right) - \int_{\cC} d\tau\, \bar{B}^{\mu\nu}(\tau) \frac{\delta^{(4)}(x-z(\tau))}{\sqrt{-g}}\right]\,.
$$
In addition, one can use (\ref{NonFinalStructureForBandBarB}) and (\ref{DefinitionOfTheTscale}) to eliminate the coefficient $\bar{B}^{\mu\nu}$ in favor of $T^{\mu\nu}$ and $m(\tau)$, which are arguably more observable, obtaining
\begin{equation} \label{orderT}
\cO_1(T)\geq 2 \eta \Re \left( \bra{\Psi^{\orto}} \hat{T}_{\mu\nu} \ket{\Psi}\right) + T^{\mu\nu} - \int_{\cC} d\tau \, m(\tau) u^{\mu} u^{\nu} \frac{\delta^{(4)}(x-z(\tau))}{\sqrt{-g}} \,.
\end{equation}

This inequality should be interpreted as follows. Given an explicit model of quantum gravity, and within it an explicit configuration of matter fields that make up a particle, one can estimate all three quantities on the right-hand side of (\ref{orderT}), namely the off-diagonal components of the stress-energy operator, its expectation value in the dominant classical state, and the total mass of the particle, respectively. Then, the consistency condition (\ref{orderT}) gives a lower bound on the scale $\cO_1$, which represents an estimate of the error when discussing the effective equation of motion for the particle. In other words, the equation of motion can be considered to be approximately valid only across scales much larger than the $\cO_1$ scale, bounded from below by inequality (\ref{orderT}).

Finally, if one needs better precision than the scale determined by (\ref{orderT}), one should take into account the dipole term in (\ref{DefinitionOfTheTbarScale}) and rederive a more precise form of the equation of motion. Still better precision would be obtained by including the dipole term in (\ref{DefinitionOfTheTscale}), which would amount to the equation of motion in the full pole-dipole approximation, and so on.

\section{\label{SecConsequencesForWEP}Status of the weak equivalence principle}

In light of the results of section~\ref{SecGeodesicEquationInQG}, it is important to discuss the status of the equivalence principle (EP). Throughout the literature, one can find various different formulations of EP, in various flavors such as weak, medium-strong, strong, and so on (see~\cite{OkonCallender,CasolaLiberatiSonego} for a review, and~\cite{AcciolyPaszko,Chowdhury,ZychBrukner2015,BrownRead,Rosi,Longhi} for various examples). Often these formulations and flavors are interpretation-dependent, and it is not always clear whether they are mutually equivalent or not, and what are the underlying assumptions and definitions used to express them.

Needless to say, such situation is less than satisfactory~\cite{OkonCallender,CasolaLiberatiSonego}, and in order to circumvent it, in this section we opt to specify one particular definition of the weak and strong equivalence principles (WEP and SEP, respectively) and to use this definition in the rest of the text. We do not aspire to claim that our definition is either equivalent to, or in any sense better than, other definitions present in the literature, and may not even correspond to the usual usage of the terminology. But for the purpose of clarity, it is prudent to fix one definition and stick to it. Therefore, in light of the results obtained in section~\ref{SecGeodesicEquationInQG}, in this section we discuss the status of WEP defined as below.

\subsection{\label{SubSecDefinitionOfEP}Definition and flavors of the equivalence principle}

The purpose of the equivalence principle is to prescribe the coupling of matter to gravity~\cite{MisnerThorneWheeler}. Its precise formulation therefore depends on the particular choice of the gravitational and matter degrees of freedom which one uses to describe matter and gravity. For the purpose of this paper,  we assume that the classical limit of quantum gravity corresponds to general relativity, which means that in this limit the fundamental gravitational degrees of freedom give rise to a nonflat spacetime metric. Given any choice of the gravitational degrees of freedom that belong to this class, in the classical framework one can formulate the equivalence principle as a two-step recipe to couple matter to gravity (we will discuss the quantum framework in subsection~\ref{SubSecEPinQG}).

Start from the classical equation of motion for matter degrees of freedom in flat spacetime, written symbolically as
\begin{equation} \label{GeneralEOMforMatterInFlatSpacetime}
\cD_{\rm flat}[\phi,\eta_{\mu\nu}] = 0\,,
\end{equation}
where $\phi$ denotes the matter degrees of freedom, $\eta_{\mu\nu}$ is the Minkowski metric, while $\cD_{\rm flat}$ is an appropriate functional describing the equation of motion for $\phi$ in flat spacetime and is assumed to be local. Given this equation of motion, couple it to gravity as follows:
\begin{enumerate}
\item Rewrite the equation of motion in a manifestly diffeomorphism-invariant form, typically by a change of variables to a generic curvilinear coordinate system,
$$
\cD_{\rm curvilinear}[\phi,g^{(0)}_{\mu\nu}] = 0\,,
$$
where $g^{(0)}_{\mu\nu}$ is still the flat spacetime metric, appropriately transformed from $\eta_{\mu\nu}$, and similarly for $\cD_{\rm curvilinear}$.
\item Promote the curvilinear equation of motion to the equation of motion in curved spacetime by replacing the flat spacetime metric $g^{(0)}_{\mu\nu}$ with an arbitrary metric $g_{\mu\nu}$,
$$
\cD_{\rm curvilinear}[\phi,g_{\mu\nu}] = 0\,,
$$
thereby specifying the equation of motion for matter coupled to gravity.
\end{enumerate}
The first step describes the matter equation of motion from a perspective of a generic curvilinear (or ``arbitrarily accelerated'') coordinate system, reflecting the principle of \textit{general relativity}. The second step simply promotes that same equation to curved spacetime as it stands, with no additional coupling of any kind. This can be loosely formulated as a statement of \textit{local equivalence between gravity and acceleration}, which is how the EP historically got its name. Also, note that these two steps operationally correspond to the standard \textit{minimal coupling} prescription~\cite{MisnerThorneWheeler}.

It is important to stress the \textit{local} nature of EP, which manifests itself in the assumption that the initial equation of motion (\ref{GeneralEOMforMatterInFlatSpacetime}) is local, and that the EP essentially does not change it at all, at any given point in spacetime. This has one important implication --- the gravitational degrees of freedom manifest themselves only through \textit{nonlocal measurements}, as tidal effects induced by spacetime curvature. We will return to this point and comment more on it later in the text.

Depending on the further specification of the matter degrees of freedom, one can distinguish between various flavors of the EP. For example, if one talks about the mechanics of point particles, one can start from the Newton's first law of motion, which states that in the absence of any forces, a particle has a straight-line trajectory in Minkowski spacetime. According to the step 1 above, the differential equation for a straight line in a generic curvilinear coordinate system is the geodesic equation,
$$
\frac{d^2 z^{\lambda}}{d\tau^2} + \mitGamma^{\lambda}_{(0)}{}_{\mu\nu} \frac{d z^{\mu}}{d\tau} \frac{d z^{\mu}}{d\tau} = 0\,,
$$
where the index $(0)$ on the Christoffel symbol indicates that it is calculated using the metric $g_{\mu\nu}^{(0)}$, which is obtained by a curvilinear coordinate transformation from the Minkowski metric $\eta_{\mu\nu}$. Then, according to step 2, one again writes the same equation, only dropping the requirement of flat spacetime metric,
$$
\frac{d^2 z^{\lambda}}{d\tau^2} + \mitGamma^{\lambda}{}_{\mu\nu} \frac{d z^{\mu}}{d\tau} \frac{d z^{\mu}}{d\tau} = 0\,,
$$
so that this time the Christoffel symbol is calculated using an arbitrary metric $g_{\mu\nu}$, and now encodes the interaction with the gravitational degrees of freedom. So one starts from the Newton's first law of motion for a particle in the absence of the gravitational field, and ends up with a geodesic equation in the presence of the gravitational field. We define this flavor of the EP as the \textit{weak equivalence principle} (WEP).

Instead of mechanical particles, one can study matter degrees of freedom described by a field theory. For example, if one starts from the equation of motion for a single real scalar field,
$$
\left( \eta^{\mu\nu}\del_{\mu} \del_{\nu} -m \right) \phi = 0\,,
$$
according to the step 1 of the EP, one can rewrite it in a general curvilinear coordinate system as
$$
\left( g_{(0)}^{\mu\nu}\nabla_{\mu} \nabla_{\nu} -m \right) \phi = 0\,,
$$
where the Christoffel symbol inside the covariant derivative is again calculated using the flat-space metric $g_{\mu\nu}^{(0)}$. Then, according to step 2 of the EP, this equation is promoted to curved spacetime as it stands, leading to
$$
\left( g^{\mu\nu}\nabla_{\mu} \nabla_{\nu} -m \right) \phi = 0\,,
$$
where now the covariant derivative is given with respect to an arbitrary metric $g_{\mu\nu}$ describing curved spacetime. Thus one arrives to the equation of motion for a scalar field coupled to gravity. We define this flavor of the EP as the \textit{strong equivalence principle} (SEP).

So in short, WEP is a statement about mechanical systems such as particles and small bodies, while SEP is a statement about fields. We emphasize again that the above definitions may or may not correspond to what is known in other literature as WEP and SEP, depending on the particular source one compares our definitions to.  For example, one can often find a definition of WEP as a statement about equality of inertial and gravitational masses. As another example, one can also find a definition of WEP as Galileo's statement that the acceleration of a particle due to the gravitational field is independent of the particle's internal details such as mass or chemical composition, a property also called \textit{universality}, emphasizing the fact that gravitation interacts with all types of particles in the same way. For an excellent review of the various formulations and flavors of EP present in the literature, see \cite{OkonCallender}.

In relation to these alternative formulations of WEP, one should note two comments. First, while the notion of ``gravitational mass'' may be useful in the context of Newtonian theory, in frameworks such as GR it is not useful, since the source in Einstein's equations is the whole stress-energy tensor, rather than any particular mass-like parameter. This renders any definition of WEP which relies on the notion of the gravitational mass unsuitable for analysis in a fundamental QG framework. Second, one can argue (see for example~\cite{OkonCallender}) that the property of universality is implicitly present even without gravitational interaction, in the Newton's first law of motion. Namely, the first Newton's law can be formulated more precisely as follows: in the absence of any forces, a particle has a straight-line trajectory in Minkowski spacetime, \textit{regardless of its internal details such as mass or chemical composition}. The Newton's first law is never spelled out in this way in textbooks, making room for a point of view that universality has something to do with gravity or the EP. However, if one accepts our definition of WEP given above, it is more natural to say that universality is a property of Newtonian mechanics, and is merely \textit{being preserved} by the WEP when one lifts the straight-line equation of motion to curved spacetime. So from this point of view, one should arguably say that WEP is merely \textit{compatible} with universality, rather than equivalent to it.

Given all these reasons, and despite the fact that these alternative definitions of WEP may be suitable in various other contexts, they are not quite adequate for the analysis given in this paper. We therefore choose to retain our own definition of WEP, while the principles of universality and equality between inertial and gravitational mass will be called as such. They are discussed in more detail in subsection~\ref{SubSecUniversalityAndGravitationalMass}.

\subsection{\label{SubSecEPinQG}Equivalence principle and quantum theory}

Adopting the above definitions of WEP and SEP, it is important to discuss their relationship. From the perspective of the classical field theory (CFT), the notion of a particle can be introduced as a localized kink-like configuration of matter fields, described as a solution of the (usually quite complicated) matter field equations. One can then employ the apparatus of multipole formalism and describe the evolution of this kink configuration in the single pole approximation, as was discussed in section~\ref{SecGeodesicEquationInGR}. Using this method, one can recover the equation of motion for a particle in classical mechanics (CM) as an approximation of the field theory. Moreover, all this can be done before or after the application of the EP, leading to the following diagram:
$$
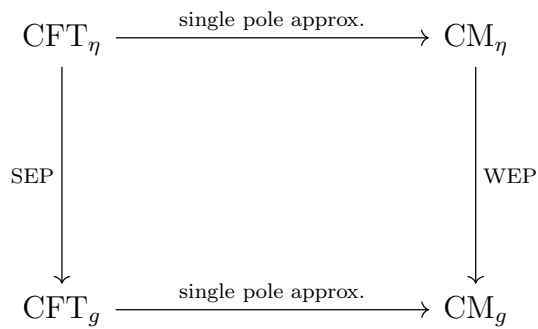
\begin{tikzcd}[row sep=7em,column sep=10em]
\text{CFT}_{\eta} \arrow[r,"\text{single pole approx.}"] \arrow[d,swap,"\text{SEP}"] &  \text{CM}_{\eta} \arrow[d,"\text{WEP}"] \\
\text{CFT}_g \arrow[r,"\text{single pole approx.}"] & \text{CM}_g 
\end{tikzcd}
$$
Here the indices $\eta$ and $g$ indicate that equations of motion in a given theory are written in flat and in curved spacetime, respectively.

The question whether this diagram commutes is nontrivial. Namely, on one hand, one can start from a flat-space classical field theory, approximate it to derive the equations of motion for a particle in flat-space classical mechanics, and then invoke WEP to reach classical mechanics coupled to gravity. On the other hand, one can first invoke SEP to couple matter to gravity at the field theory level, and then approximate it to derive the equation of motion for a particle in curved spacetime. A priori, there is no guarantee that one will reach the same equation of motion for a particle in curved spacetime using both methods.

It is in fact the existence of the local Poincar\'e symmetry that leads to the commutativity of the diagram. Namely, as was discussed in section~\ref{SecGeodesicEquationInGR}, in the curved spacetime local Poincar\'e symmetry gives rise to the covariant conservation equation for the stress-energy tensor of matter fields, and this is all one needs to reach the geodesic equation as an equation of motion for the particle, in the sense that one does not need to know the full matter field equations in curved spacetime. This establishes the $\langle \text{SEP}\to\text{single pole}\rangle $ path of the diagram. On the other hand, in flat spacetime one can also perform the calculation of section~\ref{SecGeodesicEquationInGR}, this time using the ordinary (noncovariant) conservation equation for the stress-energy tensor, which is a consequence not of the local, but rather of the global Poincar\'e invariance of Minkowski spacetime. Repeating the calculation of section~\ref{SecGeodesicEquationInGR} with the symbolic substitutions $g\to\eta$ and $\nabla\to\del$, it is not hard to conclude that one will obtain the equation of motion for a straight line in flat spacetime, again without knowing all details of the full matter field equations in flat spacetime. Then, applying WEP as discussed in subsection~\ref{SubSecDefinitionOfEP}, one reaches the geodesic equation in curved spacetime. This establishes the $\langle \text{single pole}\to\text{WEP} \rangle $ path of the diagram, concluding that the resulting equation of motion for the particle is the same in both cases, i.e., that the diagram commutes.

Let us also note that, going beyond the single pole approximation, WEP is known to be violated, with SEP remaining valid. For example, in the pole-dipole approximation, it is well known that the analogous diagram
$$
\begin{tikzcd}[row sep=7em,column sep=10em]
\text{CFT}_{\eta} \arrow[r,"\text{pole-dipole approx.}"] \arrow[d,swap,"\text{SEP}"] &  \text{CM}_{\eta} \arrow[d,"\text{WEP}"] \\
\text{CFT}_g \arrow[r,"\text{pole-dipole approx.}"] & \text{CM}_g 
\end{tikzcd}
$$
fails to commute. Namely, the $\langle \text{SEP}\to\text{pole-dipole} \rangle $ path leads to an effective equation of motion for the particle in which there is an explicit coupling of the particle's total angular momentum to the spacetime curvature~\cite{Papapetrou}. On the other hand, the $\langle \text{pole-dipole}\to\text{WEP} \rangle $ path produces the equation of motion without the curvature term. Thus, in the pole-dipole approximation, WEP fails to reproduce the correct equation of motion, since the particle is coupled to gravity in a nonminimal way, in spite of the fact that the fields which make up the particle are still minimally coupled to gravity, in line with SEP. Of course, this situation is to be expected, given that in the pole-dipole approximation the particle is no longer completely pointlike, and the coupling of the angular momentum to the curvature can be understood as a tidal effect of gravity across the ``width'' of the particle. On the other hand, one can instead argue that it would be wrong to apply WEP to the pole-dipole equation of motion for a particle. Namely, despite the fact that the latter is formally still local, it describes an object that is ``less-than-perfectly pointlike'', in the sense that its stress-energy tensor is proportional not only to a $\delta$ function but also to its derivative. From that point of view, one should not be allowed to apply the two-step prescription of EP defined above. Either way, the bottom line is that one can either declare WEP as violated or as inapplicable beyond the single pole approximation, but it cannot be declared as valid. This results in the noncommutativity of the above diagram.

Let us now turn to the quantum theory. Starting first from some quantum field theory ($\text{QFT}_{\eta}$) which describes the fundamental matter fields in Minkowski spacetime, one can take its classical limit, giving rise to some effective classical field theory ($\text{ECFT}_{\eta}$). Then, assuming that the latter features kink solutions, one can describe those using the single pole approximation, leading to classical mechanics ($\text{CM}_{\eta}$) of the corresponding particles. Finally, applying WEP one couples those particles to gravity. The resulting equation of motion will always be a geodesic equation, assuming that the initial QFT and all subsequent approximations respect the global Poincar\'e invariance of Minkowski spacetime. This symmetry guarantees the conservation of the stress-energy tensor of the matter fields throughout the sequence of approximations, leading invariably to the geodesic equation of motion for the particle.

On the other hand, it is arguably more appropriate to take an alternative, more fundamental route --- start from some fundamental quantum gravity ($\text{QG}$) model, and take the classical limit leading to some effective classical field theory ($\text{ECFT}_g$) for both matter and gravity. Then, again assuming that this theory features kink solutions, employ the single pole approximation to obtain the classical mechanics for the particle in the gravitational field ($\text{CM}_g$). Note that this is in fact precisely the program that was performed in section~\ref{SecGeodesicEquationInQG}, leading to the non-geodesic equation of motion (\ref{worldlineeq5}) for the particle. In effect, one can conclude that the following diagram fails to commute:
$$
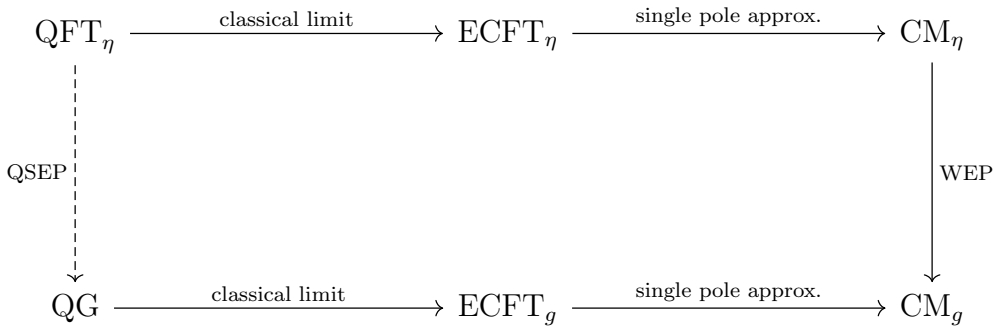
\begin{tikzcd}[row sep=7em,column sep=10em]
\text{QFT}_{\eta} \arrow[r,"\text{classical limit}"] \arrow[d,swap,dashed,"\text{QSEP}"] & \text{ECFT}_{\eta}
\arrow[r,"\text{single pole approx.}"] &  \text{CM}_{\eta} \arrow[d,"\text{WEP}"] \\
\text{QG} \arrow[r,"\text{classical limit}"] & \text{ECFT}_g \arrow[r,"\text{single pole approx.}"] & \text{CM}_g 
\end{tikzcd}
$$
As a side comment, note that the dashed $\text{QSEP}$ arrow represents some hypothetical map leading from a QFT in Minkowski spacetime to a full-blown model of QG, according to a notion that might be called a ``quantum strong equivalence principle''. It is unclear whether such a principle exists or not, let alone what its formulation is supposed to be, even if one is given precisely defined models of QFT and QG in question. We introduce it here simply for completeness, speculating that such a notion should exist, as a generalization of SEP from classical to quantum physics. It is also convenient to introduce it,  in order to close the diagram and discuss its commutativity.

It is important to stress the reason why this diagram does not commute. Recalling the details of section~\ref{SecGeodesicEquationInQG}, the local Poincar\'e symmetry is assumed to be respected at the fundamental level of QG and onwards, just like in the classical case. Moreover, the single pole approximation is used, avoiding any nonminimal coupling of the tidal forces that may be present. And yet, in spite of all that, the resulting equation of motion is not a geodesic. Looking at the equation of motion (\ref{worldlineeq5}), the reason for this is the nontrivial interference between classical states describing two classical configurations of matter, and more importantly, of gravity. In other words, the deviation from the geodesic motion is a \textit{pure quantum gravity effect} --- it is not present in the classical case, nor in the case of quantum matter in classical Minkowski spacetime. A testimony of this fact is the quantum correction term for the metric (\ref{MetricOverlap}), which features off-diagonal matrix elements of the metric operator $\hat{g}_{\mu\nu}$:
$$
\barg_{\mu\nu} = 2 \Re \left( \kappa \bra{\Psi^{\orto}} \hat{g}_{\mu\nu} \ket{\Psi} \right) + \cO(\eta) \,.
$$

In this sense, due to the noncommutativity of the above diagram, one can argue that (within the discussed framework) \textit{quantum gravity violates the weak equivalence principle}. Nevertheless, we would like to stress that our discussion regarding both strong and weak equivalence principles, based on the above prescription from subsection~\ref{SubSecDefinitionOfEP}, is inherently {\em classical}. Indeed, in steps 1 and 2 which define the implementation of EP, one considers classical equations of motion. In our case, such definition suffices, as our entangled state (\ref{eq:ent_st_1}) consists of a dominant and a sub-dominant term. Thus, we could expand our entangled equations (\ref{MetricExpansionInBeta}) and (\ref{TEIexpansionInBeta}) around the dominant classical terms, and discuss WEP in such a scenario. In fact, according to the definition of WEP, in general one can discuss its violation only {\em with respect to} some (perhaps unspecified, but assumed) classical spacetime metric. In our case, this role is played by the dominant classical metric $g_{\mu\nu}$.

In the more general case of superpositions of states which are more equally weighted, $\alpha \approx \beta$, and which consist of almost orthogonal states, $\bracket{\Psi}{\tilde{\Psi}} \approx 0$, one cannot single out a preferred classical metric, and therefore the classical definitions of SEP and WEP are inapplicable in this regime. Therefore, both equivalence principles ought to be extended to their respective \textit{quantum} domains, denoted QSEP and QWEP respectively, in the sense of the following diagram:
$$
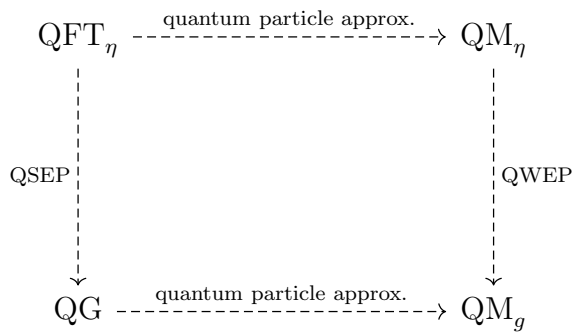
\begin{tikzcd}[row sep=7em,column sep=10em]
\text{QFT}_{\eta} \arrow[r,dashed,"\text{quantum particle approx.}"] \arrow[d,swap,dashed,"\text{QSEP}"] &  \text{QM}_{\eta} \arrow[d,dashed,"\text{QWEP}"] \\
\text{QG} \arrow[r,dashed,"\text{quantum particle approx.}"] & \text{QM}_g 
\end{tikzcd}
$$
Note that here all arrows are dashed, indicating the speculative nature of all these maps. Also, $\text{QM}_g$ represents a hypothetical theory of quantum particles coupled to a quantum gravitational field.

In this highly quantum regime ($\alpha\approx \beta$ and $\bracket{\Psi}{\tilde{\Psi}} \approx 0$), one could try to define the quantum weak equivalence principle (QWEP) in terms of the classical WEP, applied separately to each ``branch'' in the superposition. As long as the two ``branches'' $\ket{\Psi}$ and $\ket{\tilde{\Psi}}$ are themselves classical states, corresponding to the respective solutions of Einstein's equations, such a definition might seem suitable. Note that this approach is compatible with the notion of a superposed observer (see recent work~\cite{GiacominiCastroRuizBrukner} and the references therein). However, the formulation of the quantum strong and weak equivalence principles for the case of generic non-classical quantum states is an open question, outside of the scope of the current work.

Finally, the quantum version of the single pole approximation, called ``quantum particle approximation'' in the diagram above, is also not well defined --- neither conceptually nor technically. Essentially, the whole diagram represents merely a speculation about the prescriptions which ought to map between the respective theories. In addition, like in the previous cases, the commutativity of the diagram (i.e., the violation of QWEP, given the validity of QSEP) would also be an open question. In some sense, the QSEP would represent a ``true'' equivalence principle, while QWEP would be a particle-like approximate image of QSEP. Being approximate, QWEP could possibly be violated in some cases, giving rise to noncommutativity of the diagram.

\subsection{\label{SubSecUniversalityAndGravitationalMass}Universality, gravitational and inertial mass}

In light of the results of section~\ref{SecGeodesicEquationInQG}, in addition to the discussion of WEP violation, it is also important to discuss the status of the principle of universality, and the principle of equality between inertial and gravitational masses. In order to discuss them, it is instructive to study the Newtonian limit of the effective equation of motion (\ref{worldlineeq5}), as follows.

We define the Newtonian limit in the standard way~\cite{MisnerThorneWheeler} --- by assuming small spacetime curvature, nonrelativistic motion, and ignoring the backreaction of the particle on the background spacetime geometry. These approximations are implemented in the following way. First, ignoring the backreaction of the particle allows us to choose the dominant classical metric $g_{\mu\nu}$ as specified by the Newtonian line element
\begin{equation} \label{NewtonianLineElement}
ds^2 = - \left( 1 - \frac{2GM}{r} \right) dt^2 + dx^2 + dy^2 + dz^2\,,
\end{equation}
where $x^{\mu} \equiv (t,x,y,z)$ are spacetime coordinates, $M$ is the mass of the gravitational source, $r\equiv \sqrt{x^2+y^2+z^2}$, and $G\equiv l_p^2$ is Newton's gravitational constant. We will discuss the motion of a test-particle in this background, given by the effective equation of motion (\ref{worldlineeq5}). Second, the assumption of nonrelativistic motion of the particle allows us to neglect its spacelike velocity,
$$
u^k \equiv \frac{dz^k}{d\tau} \approx 0\,,
$$
leaving only the timelike component $u^0\equiv dz^0/d\tau$ nonzero (the position of the particle $z^{\mu}(\tau)$ should not be confused with the label for the third spatial coordinate $z\equiv x^3$). Finally, the assumption of small spacetime curvature allows us to neglect all terms of order $\cO(M^2)$ and higher.

Given this setup, one can easily calculate all nonzero Christoffel symbols corresponding to the dominant metric, obtaining:
$$
\itGamma^0{}_{0k} = \itGamma^0{}_{k0} = \itGamma^k{}_{00} = \frac{GM}{r^3}x^k\,, \qquad k\in\{1,2,3\}\,.
$$
One can then employ them to write the time and space components for the particle's effective equation of motion (\ref{worldlineeq5}). Using (\ref{ProjectorIdentity}) and (\ref{defF}), after some straightforward algebra, the time component of the equation of motion reduces to
$$
\frac{d^2 z^0(\tau)}{d\tau^2} = 0\,,
$$
owing to the normalization condition $u^{\mu} u^{\nu} g_{\mu\nu} = -1$ and the presence of the orthogonal projector in (\ref{worldlineeq5}). Using convenient initial conditions, this equation can be integrated to make an identification between the proper time $\tau$ and the time component of the particle's parametric equation of trajectory $x^{\mu}=z^{\mu}(\tau)$ as
$$
t = z^0(\tau) = \tau \,,
$$
reflecting the notion of global universal time of Newtonian theory. Using this result, one can show that the space components of the particle's equation of motion obtain the following form (note that the spacelike indices can be raised and lowered at will, since the spatial part of the metric (\ref{NewtonianLineElement}) is a unit matrix):
\begin{equation} \label{SpaceComponentOfEquationOfMotion}
\frac{d^2 z^k}{d\tau^2} + \frac{GM}{r^3} z^k + \eta \left[ \del_0 \barg_{0k} - \frac{1}{2} \del_k \barg_{00} - \frac{GM}{r^3} z^j \barg_{jk} \right] = 0\,.
\end{equation}
Note that here $r$ has been evaluated at the position of the particle, $r = \sqrt{z^kz_k}$, and similarly for the gradients of $\barg_{0k}$ and $\barg_{00}$. The first two terms in the equation come from the classical geodesic part $\nabla u^k$ in (\ref{worldlineeq5}), while the third term is the quantum correction, coming from the effective force term $\eta u^{\nu}u^{\sigma}  F_{\orto}^k{}_{\nu\sigma}$.

The most important aspect of equation (\ref{SpaceComponentOfEquationOfMotion}) is the similarity between the second term of the classical part and the final term of the quantum correction. The spacelike components $\barg_{jk}$ can be separated into the trace and traceless part,
$$
\barg_{jk} \equiv \frac{1}{3} h^i{}_i \, \delta_{jk} + \tilde{\barg}_{ij}\,, \qquad \tilde{\barg}^k{}_{k} \equiv 0\,,
$$
and the trace can be grouped together with the classical term, giving
\begin{equation} \label{LawOfAcceleration}
\frac{d^2 z^k}{d\tau^2} + \frac{GM}{r^3} z^k \left( 1-\frac{1}{3} \eta \barg^i{}_i \right) + \eta \left[ \del_0 \barg_{0k} - \frac{1}{2} \del_k \barg_{00} - \frac{GM}{r^3} z^j \tilde{\barg}_{jk} \right] = 0\,.
\end{equation}
Finally, multiplying the whole equation (\ref{LawOfAcceleration}) with an arbitrary positive number, called the particle's \textit{inertial mass} and denoted $m_I$, it takes the form of the Newton's second law of motion,
\begin{equation} \label{NewtonsSecondLawOfMotion}
m_I \frac{d^2 z^k}{d\tau^2} = - m_I \left( 1-\frac{1}{3} \eta \barg^i{}_i \right) \frac{G M}{r^3} z^k - \eta m_I \left[ \del_0 \barg_{0k} - \frac{1}{2} \del_k \barg_{00} - \frac{GM}{r^3} z^j \tilde{\barg}_{jk} \right] \,.
\end{equation}
One can recognize two force terms on the right-hand side. The second term is of purely quantum origin, and represents the effective force acting on the particle ultimately due to the presence of the quantum state $\ket{\tilde{\Psi}}$ in (\ref{eq:ent_st_1}). It has a non-Newtonian form, in the sense that none of its parts can be grouped together with the first force term, as was done with the trace part. The first force term, however, can be recognized as the classical Newton's gravitational force law, provided that one defines the ratio between the \textit{gravitational mass} $m_G$ and the \textit{inertial mass} $m_I$ of the particle as
\begin{equation} \label{GravitationalMass}
\frac{m_G}{m_I} \equiv \left( 1-\frac{1}{3} \eta \barg^i{}_i \right)\,.
\end{equation}

At this point we are ready to discuss the principles of universality and of the equality between gravitational and inertial masses. To begin with, it is obvious from (\ref{GravitationalMass}) that the gravitational mass is equal to the inertial mass only up to a quantum correction term. This term contains the trace of spatial components of the metric interference tensor $\barg_{\mu\nu}$, defined by equation (\ref{MetricOverlap}), from which we obtain: 
\begin{equation} \label{SpatialTraceOfMetricOverlap}
\barg^i{}_i = 2 \delta^{ij} \Re \left( \kappa \bra{\Psi^{\orto}} \hat{g}_{ij} \ket{\Psi} \right) + \cO(\eta) \vphantom{\ds\frac{A}{B_B}}\, . 
\end{equation}
It is crucial to notice that, in addition to the dependence of the off-diagonal matrix element of the metric operator, this expression also depends on the matter fields (which are present in $\ket{\Psi}$ and $\ket{\Psi^{\orto}}$), including the particle itself. Therefore, the term in the parentheses in (\ref{GravitationalMass}) cannot be reabsorbed into the constants $G$ and $M$, since these describe the external source of gravity which should remain independent of the properties of the test particle. Thus, the only possibility to cast the first force term in (\ref{NewtonsSecondLawOfMotion}) into the form of the Newton's law of gravitation, is to define the ratio between the gravitational and the inertial mass as in (\ref{GravitationalMass}). As a consequence, the principle of equality between gravitational and inertial masses is violated by the presence of the correction term coming from quantum gravity.

A similar argument can be made regarding the principle of universality. One may cancel away the inertial mass from the Newton's law (\ref{NewtonsSecondLawOfMotion}), returning to (\ref{LawOfAcceleration}) which describes the acceleration of the particle in the presence of an external gravitational field. Again, the presence of (\ref{SpatialTraceOfMetricOverlap}) in the classical gravitational acceleration term guarantees that this term depends not only on the external gravitational source, but also on the structure of the test particle itself. Moreover, the remaining quantum correction terms also depend on $\barg_{\mu\nu}$, and therefore they too carry information about the internal structure of the particle. In this sense, test particles described by different matter configurations may therefore display different accelerations, given the same background gravitational field. This means that the principle of universality is violated by the presence of the quantum gravity correction terms.

As a final comment, we should also note that $m_I$ (and consequently $m_G$ as well) is a completely free parameter in the Newtonian setup, and should be determined by the interactions of nongravitational type. In particular, the Newtonian framework does not allow us to connect $m_I,m_G$ with the effective mass $m$ of the particle, discussed in the context of (\ref{NonFinalStructureForBandBarB}) and (\ref{MassNonConservation}). This is because the total rest-energy of a particle is an inherently relativistic concept, not defined in Newtonian mechanics. On the other hand, if one goes to the relativistic framework, the notions of inertial and gravitational masses become ill-defined, since gravitational interaction cannot be described anymore by a mere force law in the Newtonian sense. Therefore, the relationship between $m$ on one side, and both $m_I$, $m_G$ on the other side, remains undefined.

\section{\label{SecConclusions}Conclusions}

\subsection{Summary of the results}

In this paper, we have discussed the effective motion of a point particle within the framework of quantum gravity, in particular the case where both matter and gravity are in a quantum superposition of the Schr\"odinger cat type. In section~\ref{SecGeodesicEquationInGR} we gave a recapitulation of the results of the classical theory, introducing the multipole formalism framework and illustrating the derivation of the geodesic equation for the motion of a particle in GR. Section~\ref{SecGeodesicEquationInQG} was devoted to the generalization of these results to the realm of the full quantum gravity. In subsection~\ref{SubSecPrelimiaries} we introduced the abstract quantum gravity framework, discussed the model of the superposition of two classical states, and established the main assumptions for the derivation of the effective equation of motion. In subsection~\ref{SubSecCovariantConservationEquation} we have analyzed in detail the quantum version of the equation for the covariant conservation of stress-energy tensor, which is a crucial ingredient in the derivation of the effective equation of motion. The explicit derivation of the equation of motion itself was then given in subsection~\ref{SubSecEquationOfMotion}, giving rise to the main results of the paper --- the equation for the stress-energy kernel (\ref{NonFinalStructureForBandBarB}), the equation for the time-evolution of the particle's mass (\ref{MassNonConservation}), and the effective equation of motion for the particle (\ref{worldlineeq5}). Most importantly, the effective equation of motion turns out to contain a non-geodesic term, giving rise to an effective force acting on the particle, as a consequence of the interference terms between the two classical states of the gravity-matter system. The last subsection~\ref{SubSecConsistency} discusses the self-consistency of the assumptions used in the above analysis, giving rise to the equation (\ref{orderT}) for the error estimate of the single pole approximation scale.

In light of the nongeodesic motion established in section~\ref{SecGeodesicEquationInQG}, it is important to discuss it in the context of the equivalence principle. This topic was taken up in section~\ref{SecConsequencesForWEP}. After we have defined various flavors of the equivalence principle in subsection~\ref{SubSecDefinitionOfEP}, the main analysis was presented in subsection~\ref{SubSecEPinQG}, discussing a possible violation of (various forms of) the weak equivalence principle, as a consequence of the nongeodesic correction to the equation of motion (\ref{worldlineeq5}). Also, given the inherently classical nature of the equivalence principle, we have also speculated on possible generalizations to the quantum realm, introducing the notions of the quantum strong and weak equivalence principles, albeit without giving explicit statements about their definitions. Finally, in subsection~\ref{SubSecUniversalityAndGravitationalMass} we have discussed the notions of universality and equality between inertial and gravitational masses in the context of quantum gravity, by studying the Newtonian limit of the equation of motion (\ref{worldlineeq5}). This analysis gave a clear interpretation that both universality and the equality between gravitational and inertial masses are violated in our context, corroborating the conclusions of the abstract analysis of the EP given in subsection~\ref{SubSecEPinQG}.

\subsection{Discussion of the results}

By far the most interesting topic to discuss in the context of the equation of motion (\ref{worldlineeq5}) is how to estimate the magnitude of the nongeodesic term. As far as the analysis of this paper goes, we can only say that this term is very small, given that it is proportional to $\eta$, which is in turn bounded from above by phenomenological argument that we do not observe superpositions of the gravitational field in nature. However, aside from this qualitative argument, in order to estimate the actual magnitude of the nongeodesic term one would need to go beyond the abstract quantum gravity formalism, and construct an explicit quantum gravity model coupled to matter fields, find some explicit kink solutions of the matter sector, and then calculate the overlap terms and the off-diagonal interference terms of the metric operator. Of course, any estimate obtained in such a way would be model-dependent. We consider this to be a feature of the abstract quantum gravity approach, since the magnitude of the nongeodesic term represents one way to operationally distinguish between different QG models. In other words, one could use equation (\ref{worldlineeq5}) to experimentally test and compare these models, at least in principle. Probably the most obvious such test would employ equation (\ref{GravitationalMass}) which relates the gravitational and inertial mass of the particle.

One result that was not discussed in detail is the nonconservation law for the effective mass of the particle, (\ref{MassNonConservation}). However, it is not really surprising that the particle's total rest energy fails to be constant in the presence of gravity-matter entanglement. As (\ref{MassNonConservation}) tells us, the nonconservation is actually a consequence of the additional effective force, which is itself a consequence of the quantum interference between two classical geometries and matter states. Nevertheless, it would indeed be interesting to study the mass nonconservation in more detail.

It is also important to discuss the generalization of our results from the case of the superposition of two classical states to many classical states. In particular, one could discuss the case where the state $\ket{\tilde{\Psi}}$ in (\ref{eq:ent_st_1}) is not a single classical state, but a superposition of many classical states,
$$
\ket{\tilde{\Psi}} = \sum_i \gamma_i \ket{\Psi_i}\,.
$$
As long as we maintain the assumption that the fidelity $F(\ket{\Psi},\ket{\PsiEnt}) \approx 1$, it is straightforward to see that all our results and conclusions still hold in the generic case. Therefore, there is no substantial difference in the analysis of a state which is a superposition of two classical states, compared to the analysis of a superposition of many classical states, as long as one of them is dominant while all others are sub-dominant. Note that in this case, even when $\beta$ is finite and $\epsilon \to 0$, the role of the metrics generated by $\ket{\Psi}$ and $\ket{\tilde{\Psi}}$ cannot be exchanged anymore, as the latter generically does not satisfy Einstein's equations. This fixes the choice of $\ket{\Psi}$ as the dominant state. A detailed quantitative description is technically more complicated, but qualitatively all results will hold for both types of states.

\subsection{Future lines of research}

One of the main lines of future work would be to perform a similar analysis as was done in this paper, but keeping the $\eta^2$ terms. This would naturally include the sub-dominant effective metric and stress-energy (\ref{gTdefTilde}), giving qualitatively new insight into the notion of quantum superpositions of two classical geometries. That analysis might provide clues about the properties of quantum gravity which could arguably hold even in the equal-weight superpositions of two classical states, defined by the choice $\alpha \approx \beta\approx 1/\sqrt{2}$ in (\ref{eq:ent_st_1}).

Alternatively, one could repeat the analysis of this paper, but in a pole-dipole approximation. This would also lead to novel effects, one of which might be a coupling of various quantum interference terms to the spacetime curvature and the angular momentum of the particle, generalizing the classical pole-dipole equation of motion \cite{Papapetrou}.

Also, given that the multipole formalism is also applicable to Riemann-Cartan spacetimes \cite{YasskinStoeger,ShirafujiNomuraHayashiOne,ShirafujiNomuraHayashiTwo,VasilicVojinovicDirakovaCestica}, the analysis of this paper could be generalized to include coupling of quantum interference terms to spacetime torsion and the spin of the particle.

Finally, one could further discuss a more general setup in which the off-diagonal terms in the covariant conservation equation (\ref{QuantumCovariantConservationTEIrewritten}) are not ignored, in the sense of going beyond the approximations (\ref{RequirementOnStressEnergyUncertainty}) and (\ref{RequirementOnMetricUncertainty}).

In addition to all of the above, one important line of research would be to study possible connections to experiments. First, one should study the counterpart of the so-called \textit{geodesic deviation equation}. Namely, in GR, the geodesic motion as such is not observable, as a consequence of the equivalence principle. As we have emphasized in subsection~\ref{SubSecDefinitionOfEP}, the EP dictates that the only way to observe gravitational degrees of freedom is via \textit{nonlocal measurements}, which are not encoded in the geodesic equation. Therefore, what one can actually observe is the change in the relative separation of two nearby geodesic trajectories, due to the tidal effects. This is in turn described by the geodesic deviation equation, which explicitly features the Riemann curvature tensor. In our case, the equation of motion (\ref{worldlineeq5}) is not a geodesic, but is still local in character, in the sense that it does contain gravitational degrees of freedom at the given point, but still it does not combine gravitational degrees of freedom of two or more points. Thus, one ought to compare the trajectories of two nearby particles, both following a trajectory determined by (\ref{worldlineeq5}). The equation governing the separation between two particles in such a setup would be a counterpart to the geodesic deviation equation of GR with a corresponding quantum correction term. It should be derived and studied in detail, in order to better understand what effects could be in principle directly experimentally observable.

Second, one could also test our results by measuring the violation of the universality and of the equality of the gravitational and the inertial mass in the semiclassical Newtonian limit.

The above list of possible topics for further research is of course not exhaustive --- one can probably study various additional aspects and topics related to this work, in particular giving more precise meaning to the notions of the quantum strong and weak equivalence principles.

\bigskip

\centerline{\bf Acknowledgments}

\bigskip

NP acknowledges the support of SQIG (Security and Quantum Information Group), the Instituto de Telecomunica\c{c}\~oes (IT) Research Unit, Ref. UID/EEA/50008/2019, funded by Funda\c{c}\~ao para a Ci\^encia e Tecnologia (FCT), project H2020 SPARTA, and the FCT projects Confident PTDC/EEI-CTP/4503/2014, QuantumMining POCI-01-0145-FEDER-031826\ \ and Predict PTDC/CCI-CIF/ 29877/2017, supported by the European Regional Development Fund (FEDER) through the Competitiveness and Internationalization Operational Programme (COMPETE 2020), and by the Regional Operational Program of Lisbon, as well as the bilateral scientific cooperation between Portugal and Serbia through the project ``Noise and measurement errors in multi-party quantum security protocols'', no. 451-03-01765/2014-09/04 supported by the Foundation for Science and Technology (FCT), Portugal, and the Ministry of Education, Science and Technological Development of the Republic of Serbia. FCT grant CEECIND/04594/2017 is also acknowledged.

MV was supported by the project ON171031 of the Ministry of Education, Science and Technological Development of the Republic of Serbia, by the bilateral scientific cooperation between Portugal and Serbia through the project ``Quantum Gravity and Quantum Integrable Models --- 2015-2016'', no. 451-03-01765/2014-09/24 supported by the Foundation for Science and Technology (FCT), Portugal, and the Ministry of Education, Science and Technological Development of the Republic of Serbia, by the bilateral scientific cooperation between Austria and Serbia through the project ``Causality in Quantum Mechanics and Quantum Gravity --- 2018-2019'', no. 451-03-02141/2017-09/02 supported by the Austrian Academy of Sciences (\"OAW), Austria, and the Ministry of Education, Science and Technological Development of the Republic of Serbia, and by the ``JESH Incoming 2016'' grant of the Austrian Academy of Sciences (\"OAW).

NP and MV would also like to thank the Erwin Schr\"odinger International Institute for Mathematics and Physics (ESI Vienna), for the warm hospitality during the workshop ``Quantum Physics and Gravity'' (29. May - 30. June 2017) and partial support.

\appendix

\section{\label{AppMultipoleFormalism}Short review of the multipole formalism}

In this Appendix we give a short review of the multipole formalism, providing some basic motivation for its introduction and a few elementary properties. A more rigorous treatment and more details can be found in~\cite{VasilicVojinovicJHEP}.

The multipole formalism revolves around the idea of expanding a function into a series of derivatives of the Dirac $\delta$ function, or $\delta$ series for short. Perhaps the easiest way to understand the $\delta$ series is to introduce it as a Fourier transform of a power series. For example, given a real-valued function $f(x)$, one can write it as a Fourier transform of $\tilde{f}(k)$ as
\begin{equation} \label{FourierTransformOfAfunction}
f(x) = \int_{\realni} dk\, \tilde{f}(k) \, e^{ikx}\,.
\end{equation}
In principle, we can expand $\tilde{f}(k)$ into power series as
$$
\tilde{f}(k) = \sum_{n=0}^{\infty} c_n k^n\,,
$$
where $c_n$ are some coefficients, substitute the expansion back into (\ref{FourierTransformOfAfunction}), and integrate term by term. Using the identity
$$
k^n e^{ikx} = (-i)^n \frac{\del^n}{\del x^n} e^{ikx}
$$
and the integral representation of the Dirac $\delta$ function
$$
\delta(x) = \frac{1}{2\pi} \int_{\realni} dk \, e^{ikx}\,,
$$
we obtain
$$
f(x) = \sum_{n=0}^{\infty} c_n \int_{\realni} dk\,  k^n \, e^{ikx} = \sum_{n=0}^{\infty} (-i)^n c_n \frac{d^n}{dx^n} \int_{\realni} dk\, e^{ikx} =  \sum_{n=0}^{\infty} 2\pi (-i)^n c_n \frac{d^n}{dx^n} \delta(x) \equiv \sum_{n=0}^{\infty} b_n \frac{d^n}{dx^n} \delta(x)\,.
$$
In the last step, we have merely renamed the coefficients in the expansion.

The above example is the most elementary construction of the $\delta$ series, providing some intuition. It is straightforward to see that one can generalize the procedure to perform the expansion around an arbitrary point $z$ instead of zero, such that
$$
f(x) = \sum_{n=0}^{\infty} b_n \frac{d^n}{dx^n} \delta(x-z)\,.
$$
The coefficients $b_n$ can be evaluated using the inverse formula,
\begin{equation} \label{DeltaSeriesCoefficients}
b_n = \frac{(-1)^n}{n!} \int_{\realni} dx\, (x-z)^n f(x)\,,
\end{equation}
and are usually called $n$-\textit{th order moments} of the function $f(x)$.
From (\ref{DeltaSeriesCoefficients}) one sees that the $\delta$ series is well defined for every function $f(x)$, which falls off to zero faster than any power of $x$ at both infinities.

Let us study an instructive example. Let the function $f(x)$ be an ordinary Gaussian, peaked around the point $x_0$,
$$
f(x) = \frac{1}{\sqrt{\pi}} e^{-(x-x_0)^2}\,.
$$
One can evaluate the coefficients in the corresponding $\delta$ series using (\ref{DeltaSeriesCoefficients}) to obtain:
$$
b_n = \left\{
\begin{array}{ccl}
\ds \sum_{k=0}^{n/2} \frac{(z-x_0)^{n-2k}}{4^k \, k!\, (n-2k)!} & & \text{ for even }n \,, \\
 & & \\
\ds - \sum_{k=0}^{(n-1)/2} \frac{(z-x_0)^{n-2k}}{4^k \, k!\, (n-2k)!} & & \text{ for odd }n \,. \\
\end{array}
 \right.
$$
It is important to note the following property --- if the expansion point $z$ does not coincide with the peak of the Gaussian, $x_0$, the magnitude of the coefficients $b_n$ in general grows with $n$. For example, if $z-x_0 = 2$, we have
$$
f(x) = \delta(x-z) -2 \frac{d}{dx} \delta(x-z) +\frac{9}{4} \frac{d^2}{dx^2} \delta(x-z) -\frac{11}{6} \frac{d^3}{dx^3} \delta(x-z) +\frac{115}{96} \frac{d^4}{dx^4} \delta(x-z) + \dots
$$
However, if the expansion point coincides with the peak, $z-x_0=0$, the magnitude of the coefficients falls off as $n$ grows:
$$
f(x) = \delta(x-z) +\frac{1}{4} \frac{d^2}{dx^2} \delta(x-z) + \frac{1}{32} \frac{d^4}{dx^4} \delta(x-z) + \frac{1}{384} \frac{d^6}{dx^6} \delta(x-z) + \dots
$$
From this simple example one can infer an important property of $\delta$ series --- the coefficients $b_n$ decrease as $n$ grows, if the expansion point is near the peak of the function $f(x)$. Turning the argument around, if we require that the coefficients decrease with $n$,
$$
|b_n|>|b_{n+1}|\,, \qquad \forall n\in \prirodni_0\,,
$$
this places \textit{a restriction on the possible values of the expansion point} $z$. This is the crucial property of the $\delta$ series, and is being used to define the ``position of the particle'' which corresponds to a distribution of matter fields described by a localized function $f(x)$.

Also, assuming that the expansion point $z$ has been chosen to be near the peak of the function, the decreasing nature of the coefficients $b_n$ allows one to approximate the function $f(x)$ by a truncated series. This formalizes the intuitive idea that if one looks at some localized distribution of matter fields from ``far away'', it will look roughly as a point particle. The truncation point then quantifies the amount of ``internal structure'' that is known about $f(x)$. One can therefore study the function $f(x)$ at various approximation levels: the \textit{single pole} approximation,
$$
f(x) \sim b_0 \delta(x-z)\,,
$$
the \textit{pole-dipole} approximation,
$$
f(x) \sim b_0 \delta(x-z) + b_1 \frac{d}{dx} \delta(x-z)\,,
$$
the \textit{pole-dipole-quadrupole} approximation,
$$
f(x) \sim b_0 \delta(x-z) + b_1 \frac{d}{dx} \delta(x-z) + b_2  \frac{d^2}{dx^2} \delta(x-z) \,,
$$
and so on.

It is completely straightforward to generalize the $\delta$ series to three (or more) dimensions, with the $\delta$ series of a function $f(\vec{x})$ around the point $\vec{z}$ defined as
\begin{equation} \label{DeltaSeriesInThreeDimensions}
f(\vec{x}) = \sum_{n=0}^{\infty} b_n^{i_1\dots i_n} \frac{\del}{\del x^{i_1}} \dots \frac{\del}{\del x^{i_n}} \delta^{(3)}(\vec{x}-\vec{z})\,.
\end{equation}
Here the indices $i_1,\dots i_n$ take values $1$, $2$ and $3$, and the inverse formula for the coefficients is
\begin{equation} \label{DeltaSeriesCoefficientsHigherDim}
b_n^{i_1\dots i_n} = \frac{(-1)^n}{n!} \int_{\realni^3} d^3x\, (x^{i_1}-z^{i_1})\dots (x^{i_n}-z^{i_n}) f(\vec{x})\,.
\end{equation}
For example, in electrostatics, one can expand the charge density $\rho(\vec{x})$ localized around the point $\vec{z}=0$ as
$$
\rho(\vec{x}) = b_0 \delta^{(3)}(\vec{x}) + b_1^i \frac{\del}{\del x^i} \delta^{(3)}(\vec{x}) + \dots
$$
According to (\ref{DeltaSeriesCoefficientsHigherDim}), the coefficients are
$$
b_0 = \int_{\realni^3} d^3x\, \rho(\vec{x}) \equiv Q\,, \qquad
\vec{b}_1 = - \int_{\realni^3} d^3x\, \vec{x} \rho(\vec{x}) \equiv -\vec{p}\,,
$$
where we recognize the total charge $Q$ and the electrostatic dipole moment $\vec{p}$ of the source. Thus we have
$$
\rho(\vec{x}) = Q \delta^{(3)}(\vec{x}) -\vec{p}\cdot \nabla \delta^{(3)}(\vec{x}) + \dots
$$
Substituting the $\delta$ series expansion of $\rho(\vec{x})$ into the formula for the electrostatic potential,
$$
\varphi(\vec{r}) = \int_{\realni^3} d^3x \, \frac{\rho(\vec{x})}{|\vec{r} - \vec{x}|}\,,
$$
and evaluating the integral, one obtains the familiar expression for the multipole expansion in electrostatics~\cite{Jackson}:
$$
\varphi(\vec{r}) = \frac{Q}{|\vec{r}|} + \frac{\vec{p}\cdot\vec{r}}{|\vec{r}|^3} + \dots
$$
This example also illustrates what type of approximation is achieved with the truncation of the $\delta$ series.

Next we generalize to time-dependent functions. If the function $f(\vec{x})$ evolves in time, while remaining localized in space, one can expand it into $\delta$ series by choosing the most convenient reference point $z(t)$ at each moment of time,
\begin{equation} \label{TimePlusSpaceSplitOfTheDeltaSeries}
f(\vec{x},t) = \sum_{n=0}^{\infty} b^{i_1\dots i_n}(t) \frac{\del}{\del x^{i_1}}\dots \frac{\del}{\del x^{i_n}} \delta^{(3)}(\vec{x}-\vec{z}(t))\,,
\end{equation}
where $t\in\realni$ is a time variable, and the coefficients $b$ are now time-dependent. Then one can introduce the proper time $\tau$, and use the identity
$$
\int_{\realni} d\tau\, \delta(t-\tau) = 1
$$
to rewrite (\ref{TimePlusSpaceSplitOfTheDeltaSeries}) in a $4$-dimensional manifestly Lorentz-invariant form
\begin{equation} \label{LorentzCovariantDeltaSeries}
f(x) = \int_{\realni} d\tau \sum_{n=0}^{\infty} b^{\mu_1\dots \mu_n}(\tau) \del_{\mu_1}\dots \del_{\mu_n} \delta^{(4)}(x-z(\tau))\,,
\end{equation}
where we have relabeled $(\vec{x},t) \equiv x$, introduced $z^0(\tau)=\tau$, used shorthand notation $\del_{\mu} \equiv \del/\del x^{\mu}$, and defined $b^0 = b^{00}=b^{000}=\dots=0$, since the time derivatives do not actually appear in (\ref{TimePlusSpaceSplitOfTheDeltaSeries}). The introduction of these auxiliary timelike components of the $b$-coefficients, demanded by Lorentz invariance, gives rise to an additional gauge symmetry of the expansion coefficients, since only the ``spatial'' components carry nontrivial information about the function $f(x)$. This additional gauge symmetry is called \textit{extra symmetry 1}, and is studied in detail in~\cite{VasilicVojinovicJHEP}.

Finally, one can make one more generalization, and introduce the notion of a $\delta$ series around a $p$-brane, a $(p+1)$-dimensional submanifold living in a $D$-dimensional spacetime manifold. Namely, we have seen that one can expand a function into a $\delta$ series around a point and around a one-dimensional line (equations (\ref{DeltaSeriesInThreeDimensions}) and (\ref{LorentzCovariantDeltaSeries}), respectively). Generalizing in that direction, one can introduce the world-trajectory of a $p$-dimensional object through $D$-dimensional spacetime $\cM$, with parametric equations $x^{\mu} = z^{\mu}(\xi^a)$ describing the trajectory as a $(p+1)$-dimensional submanifold $\Sigma \subset \cM$. Here $\mu \in \{0,\dots,D-1 \}$ and $a\in\{ 0,\dots p\}$, where $x^{\mu}$ are coordinates on $\cM$ while $\xi^a$ are intrinsic coordinates on $\Sigma$. Then, given a function $f(x)$ whose support is localized near the submanifold $\Sigma$, one can write its $\delta$ series expansion around $\Sigma$ in a fully diffeomorphism- and reparametrization-invariant way as:
\begin{equation} \label{FullyGeneralDeltaSeries}
f(x) = \int_{\Sigma} d^{p+1}\xi \sqrt{-\gamma} \sum_{n=0}^{\infty} \nabla_{\mu_1} \dots \nabla_{\mu_n} \left[ B^{\mu_1\dots \mu_n}(\xi) \frac{\delta^{(D)}(x-z(\xi))}{\sqrt{-g}} \right] \,.
\end{equation}
Here $\gamma$ is the determinant of the induced metric $\gamma_{ab} = g_{\mu\nu} u_a^{\mu} u_b^{\nu}$ on $\Sigma$, where $g_{\mu\nu}$ is the metric on $\cM$ and $u_a^{\mu} \equiv \del z^{\mu}/\del \xi^a$ are the tangent vectors of $\Sigma$. Note that, in order to ensure the correct tensorial behavior, the $B$-coefficients have been moved inside the action of the covariant derivatives. Namely, despite the fact that the covariant derivatives act with respect to $x$ and $B$'s do not depend on $x$, covariant derivatives still act nontrivially on $B$'s with the connection terms. For similar reasons, the term $\sqrt{-g}$ has been introduced to combine with the $\delta$ function into a quantity which transforms as a scalar under diffeomorphisms. Its introduction amounts merely to a suitable redefinition of $B$'s and does not modify the $\delta$ series in any nontrivial way.

The fully general $\delta$ series (\ref{FullyGeneralDeltaSeries}) has been studied in detail in~\cite{VasilicVojinovicJHEP}. For the purpose of the discussion given in the main text of this paper, we are interested in the case of a particle, i.e., a $(p=0)$-brane, moving along a $1$-dimensional timelike curve $\cC$ which is a submanifold of the $(D=4)$-dimensional spacetime $\cM$. In this case, there is only one intrinsic coordinate on $\cC$, denoted $\xi^0 \equiv \tau$, only one tangent vector
$$
u_0^{\mu} \equiv \frac{\del z^{\mu}(\xi)}{\del \xi^0} = \frac{dz^{\mu}(\tau)}{d\tau} = u^{\mu}\,,
$$
while the induced metric tensor is a $1\times 1$ matrix $\gamma_{00} = g_{\mu\nu} u_0^{\mu} u_0^{\nu}$. The parametrization of the curve $\cC$ with the coordinate $\tau$ can be chosen to fix the reparametrization gauge symmetry via the gauge-fixing condition $\gamma_{00}= -1$, which is actually the natural normalization of the tangent vector, $g_{\mu\nu} u^{\mu} u^{\nu} = -1$. Finally, one can then apply the $\delta$ series expansion (\ref{FullyGeneralDeltaSeries}) to the stress-energy tensor $T^{\mu\nu}(x)$ of the matter fields as
\begin{equation} 
T^{\mu\nu}(x) = \int_{\cC} d\tau \sum_{n=0}^{\infty} \nabla_{\rho_1} \dots \nabla_{\rho_n} \left[ B^{\mu\nu\rho_1\dots \rho_n}(\tau) \frac{\delta^{(D)}(x-z(\tau))}{\sqrt{-g}} \right] \,. \nonumber
\end{equation}
Note that the coefficients $B$ now carry two additional indices inherited from the stress-energy tensor. In the single pole approximation, one drops all terms in the sum except the $n=0$ term, truncating the series to the form
$$
T^{\mu\nu}(x) = \int_{\cC} d\tau\, B^{\mu\nu}(\tau) \frac{\delta^{(D)}(x-z(\tau))}{\sqrt{-g}} \,,
$$
as used in the main text.

\section{\label{AppProductCoherent}Separable classical states}

As mentioned in the main text, a recent study suggests that physical states of gravity and matter are generically entangled~\cite{PaunkovicVojinovic2017}. In this Appendix, we analyze a simple, yet possibly intriguing, consequence of the assumption that the overall classical gravity-matter state can be approximated by (or indeed is) the product of the gravity and the matter classical states, $\ket{\Psi} = \ket{g}\otimes\ket{\phi}$, where $\ket{g}\in \cH_G$ and $\ket{\phi}\in \cH_M$ are classical states for the gravity and the matter sector, respectively (and analogously for $\ket{\tilde\Psi}$).

To begin with
we introduce the overlaps as follows:
$$
S_G \equiv \bracket{g}{\tilde{g}}\,, \qquad S_M \equiv \bracket{\phi}{\tilde{\phi}}\,, \qquad S \equiv \bracket{\Psi}{\tilde{\Psi}} = S_G S_M\,.
$$
Note that, since in (\ref{eq:ent_st_1}) only the relative phase between $\ket{\Psi}$ and $\ket{\tilde{\Psi}}$ is important, we can reabsorb the phases of the coefficients $\alpha$ and $\beta$ into  $\ket{\Psi}$ and $\ket{\tilde{\Psi}}$, respectively. In this way, we have $\alpha,\beta\in\realni$, while only the overlap $S$ between the two coherent states carries the information about the relative phase, and is therefore complex. Moreover, since $S$ is a product between $S_G$ and $S_M$, the phase of $S$ can be distributed between $S_G$ and $S_M$ in an arbitrary way. A convenient choice is to have the phase in the matter sector, so that $S_G\in\realni$ and $S_M\in\kompleksni$. Next, we can decompose $\ket{\tilde{g}}$ and $\ket{\tilde{\phi}}$ into parts proportional to and orthogonal to $\ket{g}$ and $\ket{\phi}$, respectively,
\begin{equation} \label{OrthogonalDecompositionForGandPhi}
\ket{\tilde{g}} = S_G \ket{g} + \epsilon_G \ket{g^{\orto}}\,, \qquad \ket{\tilde{\phi}} = S_M \ket{\phi} + \epsilon_M \ket{\phi^{\orto}}\,,
\end{equation}
where $\bracket{g}{g^{\orto}} \equiv 0$, $\bracket{\phi}{\phi^{\orto}} \equiv 0$, and
\begin{equation} 
\epsilon_G \equiv \sqrt{1-(S_G)^2}\,, \qquad \epsilon_M \equiv \sqrt{1-|S_M|^2}\,. \nonumber
\end{equation}
Note that $\epsilon_G,\epsilon_M\in\realni$. Additionally, one can use (\ref{OrthogonalDecompositionForGandPhi}) to rewrite $\ket{\tilde{\Psi}}$ into the form
\begin{equation} 
\ket{\tilde{\Psi}} = S \ket{\Psi} + \epsilon \ket{\Psi^{\orto}}\,, \nonumber
\end{equation}
where
\begin{equation} 
\epsilon=\sqrt{\epsilon_M^2+\epsilon_G^2-\epsilon_M^2\epsilon_G^2 }\,, \nonumber
\end{equation}
and
\begin{equation} \label{psiorth}
\ket{\Psi^{\orto}}=\frac{\epsilon_M S_G}{\epsilon} \ket{g}\otimes\ket{\phi^{\orto}}+\frac{\epsilon_G S_M}{\epsilon} \ket{g^{\orto}}\otimes\ket{\phi}+ \frac{\epsilon_G \epsilon_M}{\epsilon} \ket{g^{\orto}}\otimes\ket{\phi^{\orto}}\,.
\end{equation}
Note that in the cases when $S_G$ and $S_M$ are large (and consequently $\epsilon_G$ and $\epsilon_M$ are small), we can neglect the final term from (\ref{psiorth}), obtaining the Schmidt form of the ``orthogonal correction'' of the state $\ket{\tilde{\Psi}}$, with respect to $\ket{\Psi}$. It is interesting to observe that such a state is always necessarily entangled, as its entanglement entropy is always bigger than zero. In other words, to obtain a nearby classical product state of gravity and matter $\ket{\tilde\Psi}$, one has to perturb the original (classical product) state $\ket\Psi$ with an entangled state
$\ket{\Psi^{\orto}} \approx \epsilon^{-1} \epsilon_M S_G \ket{g}\otimes\ket{\phi^{\orto}}+ \epsilon^{-1} \epsilon_G S_M \ket{g^{\orto}}\otimes\ket{\phi}$.



\section{\label{AppDiscussionOfTheAngle}Phase of interference terms}

%
In this Appendix, we analyze the expressions for the expectation values of the metric and the stress-energy tensors in the entangled state~\eqref{FullEntangledState}, given by~\eqref{MetricExpansionInBetaExact} and~\eqref{TEIexpansionInBetaExact}, respectively. We show that their third, interference, terms are generically different from zero, and thus contain non-trivial contributions linear in $\eta$. Since the two terms have the same form, we will consider the case of the metric operator only. 

By writing $\kappa = |\kappa| e^{i\varphi_\kappa} = F e^{i\varphi_\kappa}$ and $\bra{\Psi^{\orto}} \hat{g}_{\mu\nu} \ket{\Psi} = |\bra{\Psi^{\orto}} \hat{g}_{\mu\nu} \ket{\Psi}| e^{i\varphi_g}$, the third term of~\eqref{MetricExpansionInBetaExact} has the form
\begin{equation}
\label{eq:third_term}
2\eta\Re \left( \kappa \bra{\Psi^{\orto}} \hat{g}_{\mu\nu} \ket{\Psi} \right) = 2\eta F |\bra{\Psi^{\orto}} \hat{g}_{\mu\nu} \ket{\Psi}|\cos(\varphi_\kappa + \varphi_g)\,.
\end{equation}
In case $\varphi_\kappa + \varphi_g = \pm \pi/2$, i.e., the interference term is zero, then for {\em any other generic} choice of $\ket{\Psi'} = e^{i\delta}\ket{\Psi}$, we have that $\varphi_\kappa' + \varphi_g' \neq \pm \pi/2$. 

Indeed, changing $\ket{\Psi} \rightarrow \ket{\Psi'} = e^{i\delta}\ket{\Psi}$ induces the change of the other classical state 
\begin{equation}
\ket{\tilde\Psi} = S\ket{\Psi} + \epsilon \ket{\Psi^\orto} \ \longrightarrow \ \ket{\tilde\Psi'} = S\ket{\Psi'} + \epsilon \ket{\Psi^\orto} = S'\ket{\Psi} + \epsilon \ket{\Psi^\orto}\, \nonumber
\end{equation}
with $S' = S e^{i\delta} = |S|e^{i(\varphi_s + \delta)}$ (where $S = |S|e^{i\varphi_s}$), but the orthogonal state $\ket{\Psi^\orto}$ {\em does not} change. Thus, the phase of the matrix element from~\eqref{eq:third_term} changes to $\varphi_g' = \varphi_g + \delta$. On the other hand, the phase of $\kappa$ changes to $\varphi_\kappa' = \varphi_\kappa + \tilde\delta$ (note that $\tilde\delta$ is a function of $\delta$, see below).
Since
\begin{equation}
\begin{array}{lcl}
	\kappa & = & \alpha + \beta |S| e^{i\varphi_s} \ = \ |\kappa| e^{i\varphi_\kappa}\,, \\
	\kappa' & = & \alpha + \beta |S| e^{i(\varphi_s + \delta)} \ = \ |\kappa | e^{i\varphi_\kappa'}\,, \nonumber
\end{array}
\end{equation}
it is obvious that for a generic choice of the parameters, i.e., in but a discrete number of points, we have $\tilde\delta = \varphi_k' - \varphi_k \neq -\delta$, obtaining $\varphi_\kappa' + \varphi_g' = (\varphi_s + \varphi_\kappa) + (\delta + \tilde\delta) = \pm \pi/2 + (\delta + \tilde\delta) \neq \pm \pi/2$.

Thus, the linear correction to~\eqref{MetricExpansionInBetaExact}, and to~\eqref{TEIexpansionInBetaExact} as well, is zero only for a discrete number of the relative phases between the classical states $\ket\Psi$ and $\ket{\tilde\Psi}$. Otherwise, it is generically non-trivial.


\bibliography{wep-violation-paper}

\end{document}